\documentclass[useAMS,usenatbib]{mn2e}

\usepackage{times}
\usepackage{natbib}
\usepackage{aas_macros}
\usepackage[pdftex]{graphicx}

\title[]{Cluster Magnetic Fields from Galactic Outflows}
\author[J. Donnert, K. Dolag, H. Lesch and E. M\"uller]{J.Donnert$^{1}$\thanks{jdonnert@mpa-garching.mpg.de}
                                 K. Dolag$^{1}$, H. Lesch$^{2}$ and E. M\"uller$^{1}$\\
                                 $1$Max Planck Institute for Astrophysics, P.O. Box 1317, D--85741 Garching, Germany\\
                                 $2$Universit\"ats-Sternwarte M\"unchen, Scheinerstr.~1, D-81679 M\"unchen, Germany\\
}
\begin{document}

\date{Accepted 2008 October 22. Received 2008 October 13; in original form 2008 August 6}

\pagerange{\pageref{firstpage}--\pageref{lastpage}} \pubyear{2008}

\maketitle

\label{firstpage}

\begin{abstract}
We performed cosmological, magneto-hydrodynamical simulations to
follow the evolution of magnetic fields in galaxy clusters, exploring the possibility
that the origin of the magnetic seed fields are galactic outflows
during the star-burst phase of galactic evolution. To do this we coupled a semi-analytical
model for magnetized galactic winds as suggested by
\citet{2006MNRAS.370..319B} to our cosmological simulation. We
find that the strength and structure of magnetic fields observed
in galaxy clusters are well reproduced for a wide range of model
parameters for the magnetized, galactic winds
and do only weakly depend on the exact magnetic structure within the
assumed galactic outflows. Although the evolution of a
primordial magnetic seed field shows no significant
differences to that of galaxy clusters
fields from previous studies, we find that the
magnetic field pollution in the diffuse medium within filaments is
below the level predicted by scenarios with pure primordial magnetic seed field. We
therefore conclude that magnetized galactic outflows and their
subsequent evolution within the intra-cluster medium can fully
account for the observed magnetic fields in galaxy clusters. Our
findings also suggest that measuring cosmological magnetic fields
in low-density environments such as filaments is much more useful
 than observing cluster magnetic fields to infer their possible origin.
\end{abstract}


\section{Introduction}

Magnetic fields have been detected in galaxy clusters by radio
observations, via the Faraday rotation signal of the
magnetized cluster atmosphere towards polarized radio sources in
or behind clusters \citep{2002ARA&A..40..319C} and from diffuse
synchrotron emission of the cluster atmosphere \citep[see][for
recent reviews]{2004IJMPD..13.1549G,2008SSRv..134...93F}. However,
our understanding of their origin is still very limited.

At present, models for the origin of seed fields can be classified
in three main groups. In the first, a magnetic field is created in
shocks through the ''Biermann battery'' effect
\citep{1997ApJ...480..481K,Ryu..1998,2001ApJ...562..233M}. A subsequent
turbulent dynamo boosts it to the field
strength observed in galaxy clusters. A second class of models invokes processes that
took place in the early universe. In general, they predict that
magnetic seed fields fill the entire volume of the universe;
however the coherence length of the field crucially depends on the
details of the models \citep[see][for a
review]{Grasso..PhysRep.2000}. Finally, galactic winds
\citep[e.g.][]{Volk&Atoyan..ApJ.2000} or AGN ejecta \citep[e.g.][
and references therein]{1997ApJ...477..560E,Furlanetto&Loeb..ApJ2001} can produce
magnetic fields and pollute the proto-cluster region. In such
models, the magnetic field can also originate from an early
population of dwarf, starburst galaxies
\citep{1999ApJ...511...56K} at relatively high redshift ($z \approx 4 - 6$).

In previous work, non radiative simulations of galaxy clusters within a cosmological
environment which follow the evolution of a primordial magnetic
seed field were performed using Smooth-Particle-Hydrodynamics
(SPH) codes
\citep{1999A&A...348..351D,2002A&A...387..383D,2005JCAP...01..009D}
as well as Adaptive Mesh Refinement (AMR) codes
\citep{2005ApJ...631L..21B,2008A&A...482L..13D,2008ApJS..174....1L}.
Although these simulations are based on different numerical
techniques they show good agreement in the predicted
properties of the magnetic fields in galaxy clusters, when the evolution of an initial
magnetic seed field is followed. This work has also  demonstrated, that
the properties of the final magnetic field in galaxy clusters do
not depend on the detailed structure of the assumed initial
magnetic field. The spatial distribution and the structure of the
predicted magnetic field in galaxy clusters is primarily
determined by the dynamics of the velocity field imprinted by
cluster formation \citep{1999A&A...348..351D,2002A&A...387..383D}
and compares well with measurements of Faraday rotation.

The creation of magnetic fields in shocks
through the ''Biermann battery effect''
\citep{1997ApJ...480..481K,Ryu..1998}, and subsequent
turbulent dynamo action can be followed as well as a prediction can be made for 
magnetic field values from velocity fields inferred in
cosmological simulations \citep{2008Sci...320..909R}. Both methods
predict magnetic field strengths in filaments with somewhat higher
values \citep[e.g. see ][]{2004PhRvD..70d3007S} than found in
simulations that follow the evolution of a primordial magnetic
seed field.

Faraday rotation can be observed in several
radio galaxies located at different radial distances with
respect to the cluster center. Motivated by 
numerical simulations \citep{2001A&A...378..777D}, the observed
magnetic field is often modelled with a radially-declining
 field strength and a power law spectral structure.
>From such observations, once can constrain the power law spectral index   
\citep{2004A&A...424..429M,2006A&A...460..425G} or directly reconstruct the 
power spectrum of the magnetic field \citep{2003A&A...412..373V,2005A&A...434...67V}. 
Given the sparse observational data available at
the moment, a degeneracy exists between the central value of the
magnetic field and its rate of radial decline 
	\citep[see for example][]{bonafede08,2008A&A...483..699G}, for which detailed predictions from simulations
can be useful in breaking the degeneracy. The simulations must therefore examine 
different possible magnetic field origins in galaxy clusters in order to test
the robustness of the inferred magnetic field properties.

Recently the validity of models that produce a cluster
magnetic field from galactic winds has been supported by a
semi-analytic modelling of galactic winds
\citep{2006MNRAS.370..319B}. However, these models are unable to predict
how the magnetic fields produced by the ejecta of the galaxies
are compressed and amplified by the process of structure
formation. Therefore, the structure of the
final magnetic field in galaxy clusters cannot yet be predicted by these models. 
In this
work we extend these studies by directly incorporating the
galactic outflow model in magneto-hydrodynamical simulations of
structure formation.

The paper is structured as follows: In section \ref{sim setup} we
present the details of the cosmological setup, concentrating especially on the
coupling of the semi-analytic model to the cosmological
simulations. Details of the wind
model and  the seed magnetic
field are presented in section \ref{wind_model} and appendix
\ref{E2B}. The general results of our simulations are presented in
section \ref{results_all} and in section \ref{results} we
discuss our findings, particulary how galaxy clusters formed in
our simulations. Finally, we present our conclusions in section
\ref{conclusions}.

\section{Simulation setup} \label{sim setup}

To investigate the evolution of a magnetic seed field produced by
galactic outflows one needs to couple the galaxies formed in a
cosmological, magneto-hydrodynamical simulation with a model for
the galactic outflows.

\subsection{Initial conditions}

\begin{figure}
\centering
\includegraphics[width=0.45\textwidth]{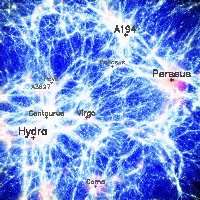}
\caption{Shown is a visualization of the constrained local
universe simulation at redshift $z=0$ using {\tt SPLOTCH}
\citep{2008arXiv0807.1742D}. Structures which can be identified
with their counterparts in the real universe are labelled. An
animation flying through the simulation can be downloaded from the MPA Website
\protect\footnotemark.
}\label{localuniverse}
\end{figure}

\footnotetext{{\tt
http://www.mpa-garching.mpg.de/galform/ \\ data\_vis/index.shtml\#movie12}} We used a constrained realization of the local
universe \citep[see][and references therein]{2005JCAP...01..009D},
with initial conditions similar to those used by 
\citet{2002MNRAS.333..739M} in their study of structure formation
in the Local Universe. The initial density fluctuations were
constructed from the IRAS 1.2-Jy galaxy survey by smoothing
the observed galaxy density field on a scale of 7 Mpc, evolving it
linearly back in time, and then using it as a Gaussian constraint
\citep{Hoffman1991} for an otherwise random realization of the
$\Lambda$CDM cosmology. The volume constrained by the
IRAS observations covers a sphere of radius $\sim 115$ Mpc centered on
the Milky Way. This region is sampled with high resolution dark
matter particles and is embedded in a periodic box of $\sim 343$
Mpc length.  The gravitational softening length used is 
$\epsilon = 14\,\mathrm{kpc}$, which corresponds to the smallest 
SPH smoothing length reached in the dense centers of halos. 
In low density regions, the resolution is lower.
The region outside the constrained volume is filled
with dark matter particles at lower resolution, allowing a good
coverage of long range gravitational tidal forces. Many
of the most prominent clusters observed locally can therefore be identified
directly with halos in the simulation, and their positions and
masses agree well with their simulated counterparts. Figure
\ref{localuniverse} shows a rendering of the simulation at
redshift $z=0$ where some of the identified structures are
labelled.

In this work we use extended initial
conditions where the original high resolution dark matter
particles are split into gas and dark matter particles with
masses of $0.69 \times 10^9\; {\rm M}_\odot$ and $4.4 \times
10^9\; {\rm M}_\odot$ respectively. The most massive clusters in
our simulations are hence resolved by nearly one million
particles. The gravitational force resolution (i.e. the gravitational 
softening length) of the simulations was set to be $14\,{\rm
kpc}$, which is comparable to the inter-particle separation
of the SPH particles in the dense centers of our simulated
galaxy clusters.

\subsection{Seeding strategy}\label{seed_strat}

\begin{figure}
\centering
\includegraphics[width=0.5\textwidth]{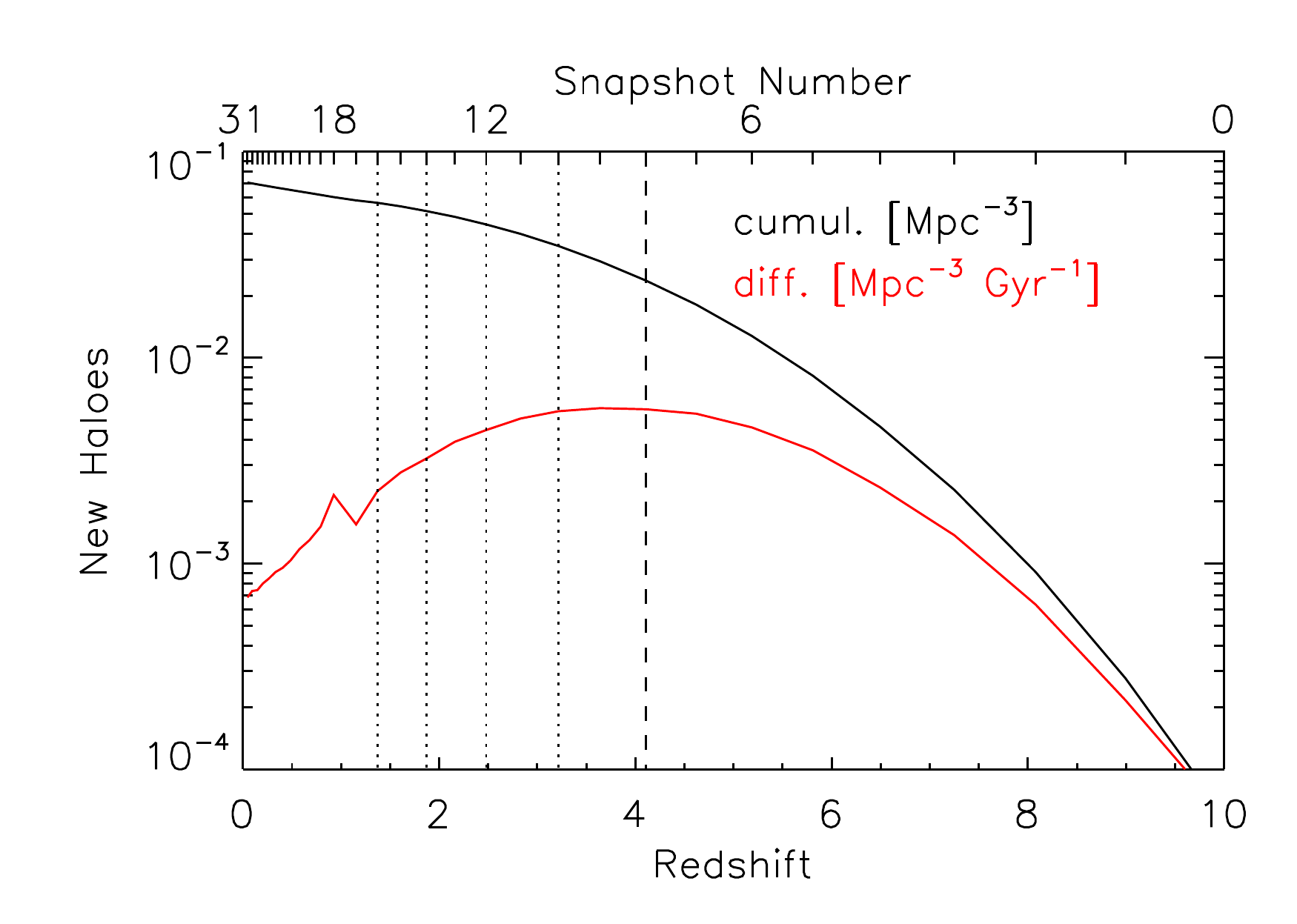}
\caption{Newly formed haloes per Mpc cubed as a function of redshift in the
reference simulation, involving a primordial magnetic seed
\citep{2005JCAP...01..009D}. The black curve shows the
cumulative distribution, the red curve the differential one (per Gyr). The dashed
line marks the time selected for the single seeding, the dotted
line marks the additional seeding times in the multi-seed
run.}\label{newhaloes}
\end{figure}

In principal one would like to follow the dynamics of galactic
outflows driven by stellar activity (either star-formation or 
star-burst) self-consistently in cosmological simulations.
However, so far this it not possible. Although cooling and
star-formation processes can be followed in standard cosmological
simulations, the detailed interaction with magnetic fields
 leads to regions in which the magnetic field pressure exceeds the thermal
one, especially in high resolution studies like those we plan to perform.
In absence of any dissipative process when performing ideal
MHD simulations, these regions then dominate the time-stepping in
the simulations and cause them to stall.
We therefore took a simpler approach by coupling a
semi-analytical recipe for star-burst-driven, magnetized outflows
\citep{2006MNRAS.370..319B} with non-radiative, cosmological MHD
simulations. In general, this should be implemented as a
continuous process, starting at very high redshift
$z\approx10$ \citep{1999ApJ...511...56K}. For practical reasons,
we approximate this process by an instantaneous magnetization of
all galaxies inferred in the simulations at a selected instance
in time (e.g. $z=4.1$), and then evolve the simulation with 
magnetic seed fields until the present day. As the cosmological
simulation evolves, the magnetized gas is stripped from its host
galaxies as it  falls towards the dense gas contained in galaxy
clusters. The intra cluster medium (ICM) is thereby 
enriched by the stripped, magnetized gas of several hundreds of
galaxies. Furthermore, the magnetized gas is processed by
shear-flows and turbulence within the cluster atmosphere, which
eventually leads to the magnetic field strength and structure
observed today in galaxy clusters. To justify our approximation we
also performed one simulation where we repeated the seeding
process several times and demonstrated that for cluster
magnetic fields these further seedings are irrelevant, as they
mainly involve newly-formed galaxies outside the proto-cluster
region. The redshift $z=4.1$ of the single seeding event was
chosen  as a compromise between complete formation of all relevant galaxies
(e.g. choosing a low seeding redshift) and
avoidance of galaxy mergers or destruction in 
proto-clusters (e.g. choosing a high seeding redshift). Figure
\ref{newhaloes} shows the number density of formed halos as a function of
redshift (differential per Gyr and cumulative) as inferred from our
reference run with a cosmological seed field
\citep{2005JCAP...01..009D}. The dashed line marks the preferred
redshift of seeding which is chosen to ensure that most of the
galaxies in the proto-cluster region are formed and not yet merged
with others, destroyed or stripped from their hot gas haloes in
the proto-cluster. The dotted line marks the additional seeding
times used in the multi-seed reference run. As seeding targets we 
choose galaxies which are newly formed (i.e. since the last
seeding) and whose mass is smaller than $M_{\mathrm{Halo}} = 3
\times 10^{12} \, M_{\odot}$ since our wind model does not apply for
group like objects. Note that the constraint on mass is only
of relevance for the low redshift seeding in the multi-seed
reference run.

\subsection{Numerical Method}\label{sims}

All simulations were performed using the MHD implementation in GADGET-2
\citep{springel2005}, similar to that used in
\citet{2005JCAP...01..009D}. Details of the MHD implementation can
be found in \citet{2008arXiv0807.3553D}. The code is based on an entropy-conserving
formulation of Smooth Particle Hydrodynamics (SPH) \citep{2002MNRAS.333..649S}. It was
supplemented with a treatment of magnetic field using ideal MHD 
\citep[see][]{2008arXiv0807.3553D}. Besides following the
induction equation for the magnetic field, we take magnetic back-reaction
into account using a symmetric formulation of the Lorentz
force based on the Maxwell tensor. The treatment of magnetic
fields in SPH was improved by explicitly subtracting the part of
the magnetic force which is proportional to the divergence of the
magnetic field, as described in \citet{2001ApJ...561...82B}. This
helps to keep numerically induced divergence of
the magnetic field at negligible values. It also helps to avoid 
instabilities of the MHD formulations in SPH, especially in
regions where magnetic field pressure substantially exceeds the
thermal pressure.

Note that in simulations without radiative cooling like  ours
the magnetic field pressure stays well below
the thermal one, even within the cores of the most massive galaxy
clusters. In very strong shocks however it can  still happen that
the magnetic field is compressed so substantially that magnetic
forces dominate the thermal ones for brief periods of time. Such
situations are handled more accurately with our new formulation,
which includes several other, small improvements in the numerical
treatment. For detailed investigations and information on the code
and its performance see \citet{2008arXiv0807.3553D}.

\begin{figure}
    \centering
    \includegraphics[width=0.5\textwidth]{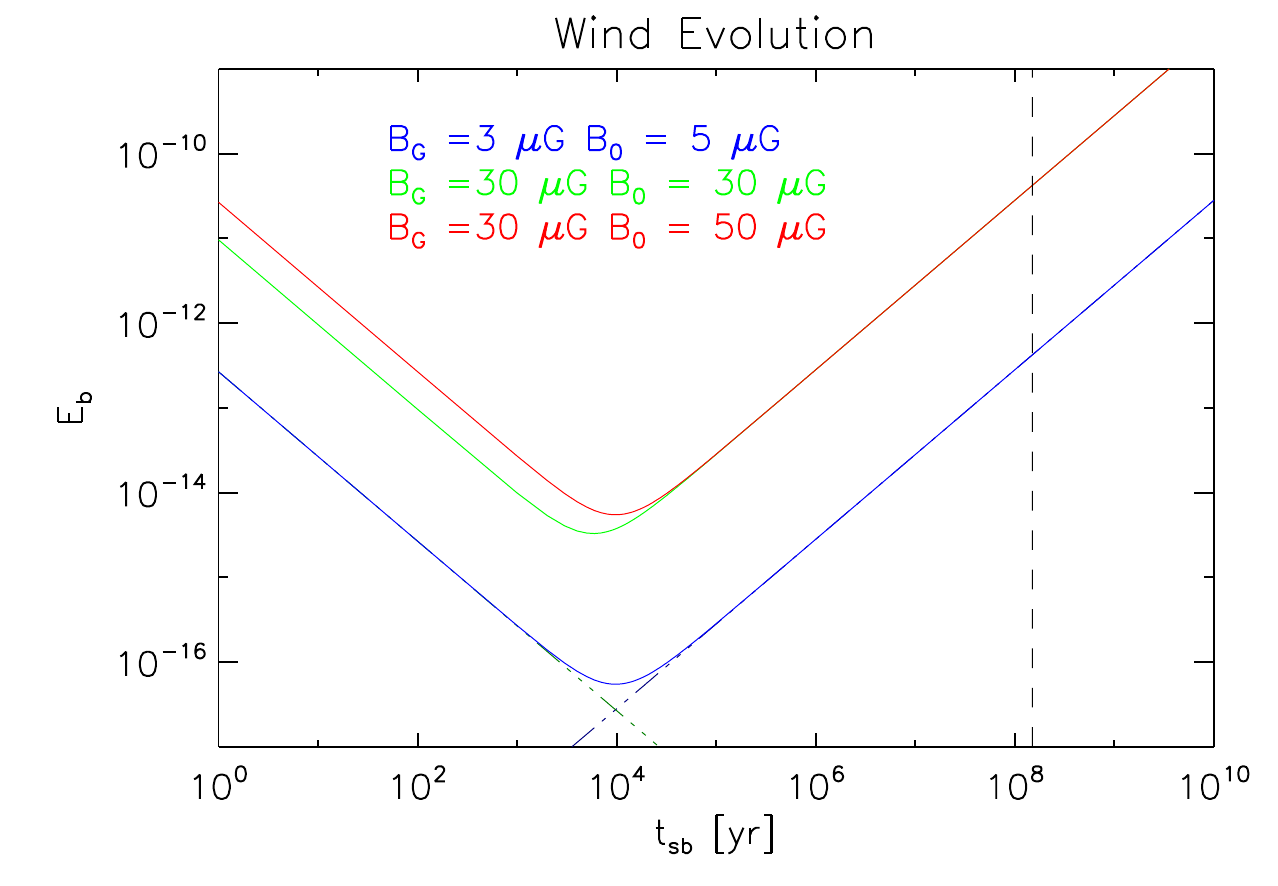}
		 \caption{Evolution of the magnetic energy in a wind bubble of a small galaxy ($M_{\mathrm{ISM}} = 10 \times 10^{12} M_{\odot}$) for
    different disc and halo parameters ($B_{\mathrm{G}},B_{0}$).
    The dashed line marks the star-burst time of
    $150\,\mathrm{Myr}$ \citep{2001astro.ph..6564D}, the
    dot-dashed lines mark the envelopes of the two wind phases.
    For the chosen $t_{\mathrm{sb}}$  the initial energy in the
    wind bubble $B_{\mathrm{0}}$ does not influence the bubble
    energy at late times.}\label{t_sb}
\end{figure}

\section{Simulations}\label{wind_model}

To identify galaxies in our simulations, we applied SUBFIND
\citep{2001MNRAS.328..726S} which uses a friends-of-friends
(FoF) algorithm to  identify locally over-dense, self-bound
particle groups associated with galaxies, even when they are
inherited within a larger parent group. Originally, this
algorithm was based on pure dark matter simulation, so we utilized
a modified version to handle the gas component as well 
(for more details see Dolag et al. 2008). \\

\subsection{The Adopted Wind Model}\label{wind_model}

\begin{table}
    \centering
    \begin{tabular}{c|c|c}
        Parameter & Value & Source \\ \hline
        $R_{0}$		     & $400\,\mathrm{pc}$ 			& \citep{1988AA...190...41K}  \\
        $B_{0}$ 		     & $ 50,5,0.5\,\mu \mathrm{G}$      & \citep{1988AA...190...41K}  \\
        $B_{\mathrm{G}}$  & $30,3,0.3\,\mu\mathrm{G}$ 	&  \citep{2004IAUS..217..436C}, \\
        						  & 								 	& \citep{2001AA...378...40S} \\
        $\dot{M}_{\star}$ & $10\,\mathrm{M}_{\odot}/{\mathrm{yr}}$ & \citep{2001astro.ph..6564D} \\
        $t_{\mathrm{sb}}$ & $ 150\,\mathrm{Myr}$      & \citep{2001astro.ph..6564D} \\
        $M_{\mathrm{ISM}}$& $<3 \times 10^{12} \mathrm{M}_{\odot}$  & from simulation
    \end{tabular}
    \caption{Summary of the parameters used for the wind model. Except
    for $B_{\mathrm{G}}$ all other parameters are based on
    observations of M82. The parameter set is scaled to the
    simulated galaxies by $M_{\mathrm{ISM}}$ }. \label{param_table}
\end{table}

To obtain the size and strength of the magnetic field in a
star-burst driven galactic outflow for each of the identified
galaxies, we adapted the semi-analytical model by
\citet{2006MNRAS.370..319B}. This model assumes an adiabatic
expansion of a spherical gas bubble with homogeneous energy density,
fed continuously by the starburst-driven outflows.
The initial bubble size (before the starburst) can
then be characterized by a field strength $B_{0}$ and a radius
$R_{0}$. The bubble expansion is driven by the starburst in the galaxy,
which is expelling gas at the rate $\dot{M}_{\star}$, dragging the
frozen-in magnetic field $B_{\mathrm{G}}$ into the ICM. The wind
velocity is a function of the star formation rate, following
\citet{2005ChJAA...5..327S} we set the ISM constant $K=0.5$, which determines the dependence of wind velocity and mass outflow rate on ISM properties (evaporation parameter and blast wave speed):
\begin{eqnarray}
v_{\mathrm{w}} &= 320 \sqrt{2}  \cdot
\left(\frac{\dot{M}_{\star}}{\mathrm{M}_{\odot}/{\mathrm{yr}}}\right)^{0.145} \, \frac{km}{s} .
\end{eqnarray}
The galaxy thus injects magnetic energy into the wind at a rate
\begin{eqnarray}
\dot{E}_{\mathrm{B}_{\mathrm{in}}}
&=\epsilon_{\mathrm{B}_{\mathrm{in}}}
\frac{\dot{M_{\mathrm{w}}}}{\bar{\rho}_{\mathrm{in}}} ,
\end{eqnarray}
where the injected mass rate is \citep{2005ChJAA...5..327S}
\begin{eqnarray}
\dot{M_{\mathrm{w}}} &= 2.5 \left(\frac{\dot{M}_{\star}}{\mathrm{M}_{\odot}/{\mathrm{yr}}}\right)^{0.71}\,
\frac{M_{\odot}}{yr} ,
\end{eqnarray}
and the average injected mass density $ \bar{\rho}_{\mathrm{in}} $ 
follows from the injected mass blown through a spherical surface at the
galactic radius, which is assumed to be a fraction of the virial
radius $R_{\mathrm{g}}=0.1\, R_{200}$:
\begin{eqnarray}
\bar{\rho}_{\mathrm{in}}  &= \frac{\dot{M}_{\mathrm{w}}}{4\pi
R_{\mathrm{g}}^{2} v_{\mathrm{w}}} .
\end{eqnarray}
The injected magnetic field energy density $\epsilon_{B_{\mathrm{in}}}$ dereases 
adiabatically
as $B^{2} \propto \rho ^{\frac{4}{3}}$ (see App.A in
\cite{2006MNRAS.370..319B}):
\begin{eqnarray}
\epsilon_{B_{\mathrm{in}}} &= \frac{B_{\mathrm{G}}^{2}}{8\pi}
\left( \frac{\bar{\rho}_{\mathrm{in}}}{\bar{\rho}_{\mathrm{ISM}}}
\right)^{\frac{4}{3}} ,
\end{eqnarray}
 where $\bar{\rho}_{\mathrm{ISM}} = M_{\mathrm{ISM}} /
  \frac{4}{3} R_{\mathrm{G}}^{3}$ is the average ISM mass
  ($M_{\mathrm{ISM}}$) density inside a sphere of galactic radius. 
Finally, the predicted injected magnetic energy will be

\begin{eqnarray}
\dot{E}_{\mathrm{B}_{\mathrm{in}}} &=
\frac{1}{2}B_{\mathrm{G}}^{2}R_{\mathrm{G}}^{2}v_{\mathrm{w}} \left(
\frac{\dot{M}_{\mathrm{w}}R_{\mathrm{G}}}{3
v_{\mathrm{w}}M_{\mathrm{ISM}}} \right)^{\frac{4}{3}} .
\end{eqnarray}
Neglecting shear amplification the time evolution
of the energy in the sphere is given by
\begin{eqnarray}
\frac{\mathrm{d}}{\mathrm{d}t}E_{\mathrm{B}}(t) &=&
\dot{E}_{\mathrm{B}_{\mathrm{in}}}(t) - \frac{1}{3}
\frac{\dot{V}_{\mathrm{w}}(t)}{V_{\mathrm{w}}(t)} E_{\mathrm{B}}(t)  \\ 
&=& \dot{E}_{\mathrm{B}_{\mathrm{in}}}(t) - \frac{1}{t}
E_{\mathrm{B}}(t) .
\end{eqnarray}
Here local alignment between magnetic field and wind direction
can not be assumed\footnote{This results in the factor $1/3$ in the
dilution term} because of turbulent instabilities and random
motions in the gas. Still global structures might develop, because
the magnetic field is dynamically unimportant and follows the
preferred wind direction, parallel to the rotation axis of the
halo. Figure \ref{t_sb} shows the evolution of the bubble energy
as a function of the effective starburst time for several different
values of the disc and halo parameters ($B_{\mathrm{G}},B_{0}$) of the
model. The model is based on a galaxy with ISM mass of
$M_{\mathrm{ISM}} = 10 \times 10^{12} M_{\odot}$. The initial bubble
energy is diluted on reasonable starburst time scales, and therefore
does not affect the final magnetic energy contained in the wind.  The
starburst time $t_{\mathrm{sb}}$ is therefore degenerate with
$B_{\mathrm{G}}$ over a wide energy range. For more details on the
wind model see \citet{2006MNRAS.370..319B}.\\ 
Given the mass-cut
on $M_{\mathrm{ISM}}$, the resulting wind velocity is sufficient to
reach every particle of a halo during the star-burst time scale
assumed. Therefore we always seed the complete halo, by setting the
magnetic field for every particle. Due to the friends-of-friends
algorithm used to identify the structures, overlapping of halos is
impossible and a minimum halo distance is defined by the linking
length.

\subsection{Applying the Wind Model}

  Contrary to \citet{2006MNRAS.370..319B}, we cannot follow (and
  integrate) the evolution of the wind model over cosmic time based on
  semi-analytical modeling. Instead we are holding the numerical
  simulation at certain epochs and identify newly formed galaxies. We
  then integrate the wind model for every of these galaxies assuming a
  generic star-formation rate and a generic star-burst time. This
  effectively mimics the magnetic seeds obtained from such a galaxy at
  these epochs. The procedure is applied (in an approximate way)
  instantaneously. Given the mass-cut on $M_{\mathrm{ISM}}$, the
  resulting wind velocity is so large that the wind can reach every
  particle within the virial radius of the halos during the assumed
  star-burst time scale. Hence, we are always seeding all particles
  within the virial radius, which by construction do not overlap.
  Continuing the magnetohydrodynamic simulation we follow the complete
  dynamics of the magnetized atmospheres, including their stripping
  and mixing in the denser environment (e.g. filaments or clusters
  atmospheres).  The galactic virial radius selected for the seeding
  procedure is typically smaller than the radius of $100\,
  \mathrm{kpc}$ inferred for the wind driven bubbles in
  \citet{2006MNRAS.370..319B}. The magnetised bubbles of the galaxy
  sized halos are stripped by ram pressure effects, which are already
  important in filaments (see \citep{2006MNRAS.370..656D},
  \citep{2006MNRAS.373..397S}, \citep{2008arXiv0808.3401D}), as well
  as partially ejected by interactions with other galaxies.  The
  stripped gas gets mixed with the cluster atmosphere, and large scale
  motions typically distribute this magnetized material across a much
  scale than the expanded bubble size predicted by the semi-analytical
  models.

The instantaneous seeding process is implemented in the simulations in
the following way:
\begin{itemize}
	\item The simulation is stopped at the seeding redshift
          (usually $z=4.2$), and the newly formed galaxies (since the
          last seeding) are identified.
	\item The gas mass of the halos identified in the simulation
          is combined with the wind model to estimate the magnetic
          energy contained in the wind (see section \ref{wind_model}).
	\item The magnetic moment of a dipole/quadrupole having the
          same magnetic energy is calculated. A smoothing length of
          $14\,\mathrm{kpc}$ is used as softening length in the
          integral (appendix \ref{E2B}).
	\item The magnetic field vector of every gas particle in the
          corresponding halo is set to match the dipole/quadrupole.
          The field of every particle inside the softening length is set to a
          random orientation and the field strength at the softening length.
         \item The simulation is continued with the newly added
           magnetic field.
\end{itemize}

The parameters of the model can in principle be inferred from
observations of individual galaxies. Table \ref{param_table} shows
some observational constraints on the parameters as obtained from one
of the best observed, starbursting galaxies, M82.  Note that in
starbursting galaxies usually disc fields up to $100\,\mu\mathrm{G}$
are  observed (R.Beck, priv.com.). This is the case on scales
smaller than the resolution archieved by our simulations
($14\,\mathrm{kpc}$). We rather follow  the conservative approach in
\citet{2006MNRAS.370..319B} using a field of several tens of $\mu
\mathrm{G}$.\\ 
M82's observed mass of $6\times10^{9} \,\mathrm{M}_{\odot}$
\citep{1998PASJ...50..227S} is smaller than the average mass of the
seeded halos (table \ref{mseed_stats}). To take this into account, we
assume that the mass of the ISM $M_{\mathrm{ISM}}$ in the model is
equal to that of the halo mass in the simulation, and not from M82.
There is further mass dependence in the mass outflow rate
$\dot{\mathrm{M}}_{\star}$. Therefore the value inferred from M82
might be underestimated and the obtained wind energy represents a
lower limit for large halos. \\ As the values quoted are still quite
uncertain, we performed simulations where we varied the value for the
halo and disc magnetic field ($B_{0}, B_{G}$) within a wide range.

The magnetic field structure within galactic outflows and the
properties of the resulting wind driven bubble are largely
unconstrained as well.  In M82, \citet{1994AA...282..724R} observe a
symmetric poloidal magnetic field structure in the core, while in
edge-on galaxies combinations of symmetric (S0) and antisymmetric (A0)
field structures are observed (R.Beck, priv.com.). A0 and S0 dynamo
modes can be approximated by a softened dipole and quadrupole field
structure, respectively. We therefore describe the structure of the
magnetic seed field as a (softened) dipole field, which we normalized
so that the field energy corresponds to the energy in the magnetic
bubble inferred from our semi-analytical model. The orientation of the
dipole is chosen to align with the spin of the underlying dark matter
halo. To verify the dependence of our results on the detailed
structure of the seed fields we also performed a run where we used a
quadrupole instead of a dipole structure for the magnetic seed
field. \\

\begin{table}
\centering
\begin{tabular}{c|c|c|c}
\hline
Snapshot & Redshift & $N_\mathrm{halos}$ & $<M_\mathrm{halos}>$ $[M_\odot]$ \\
\hline
$8$  & $4.1$ & $24731$ & $1.2\times 10^{11}$ \\
$10$ & $3.2$ & $10483$ & $6.9\times 10^{10}$ \\
$12$ & $2.5$ & $12467$ & $6.7\times 10^{10}$ \\
$14$ & $1.9$ & $11326$ & $6.6\times 10^{10}$ \\
$16$ & $1.4$ & $9838$  & $6.6\times 10^{10}$ \\
\hline
\end{tabular}
\caption{Halo statistics of the multiple seed run. The rows give
the output number, the corresponding redshift, the number of new
haloes identified since last seeding and the mean mass of these
haloes, respectively.}\label{mseed_stats}
\end{table}

Including the reference run ({\it Control Run}) which follows the
evolution of a cosmological seed field, we performed six different
simulations to explore the parameter space. In all runs the starburst
time $t_{\mathrm{sb}}$ is kept fixed at $150$ Myr.  We performed three
simulations for different values of the magnetic fields (halo $B_{0}$
and disc $B_{\mathrm{G}}$ ) which all have a dipole like
structure. Starting from the M82 like value of $50\mu \mathrm{G}$
({\it Dipole}), we also performed a run with one tenth of the value
({\it 0.1 Dipole}) and one hundredth ({\it 0.01 Dipole}). We used the
{\it Dipole} parameter set from above to perform a run with a
quadrupole structure ({\it Quadrupole}). We further repeated the {\it
  0.1 Dipole} run and performed four additional seeding episodes ({\it
  Multi Seed}) to all newly formed haloes as summarized in table
\ref{mseed_stats}. Form the results (e.g. compare left and right panel
in the middle row of figure \ref{vis30}) its clear that after the
first seeding, many new haloes are still seeded, but they are
typically smaller in mass (see last column in table \ref{mseed_stats})
and preferentially located outside the proto-cluster region.

\begin{figure*}
    \centering
    \begin{minipage}[t]{0.49\textwidth}
        \includegraphics[width=1.0\textwidth]{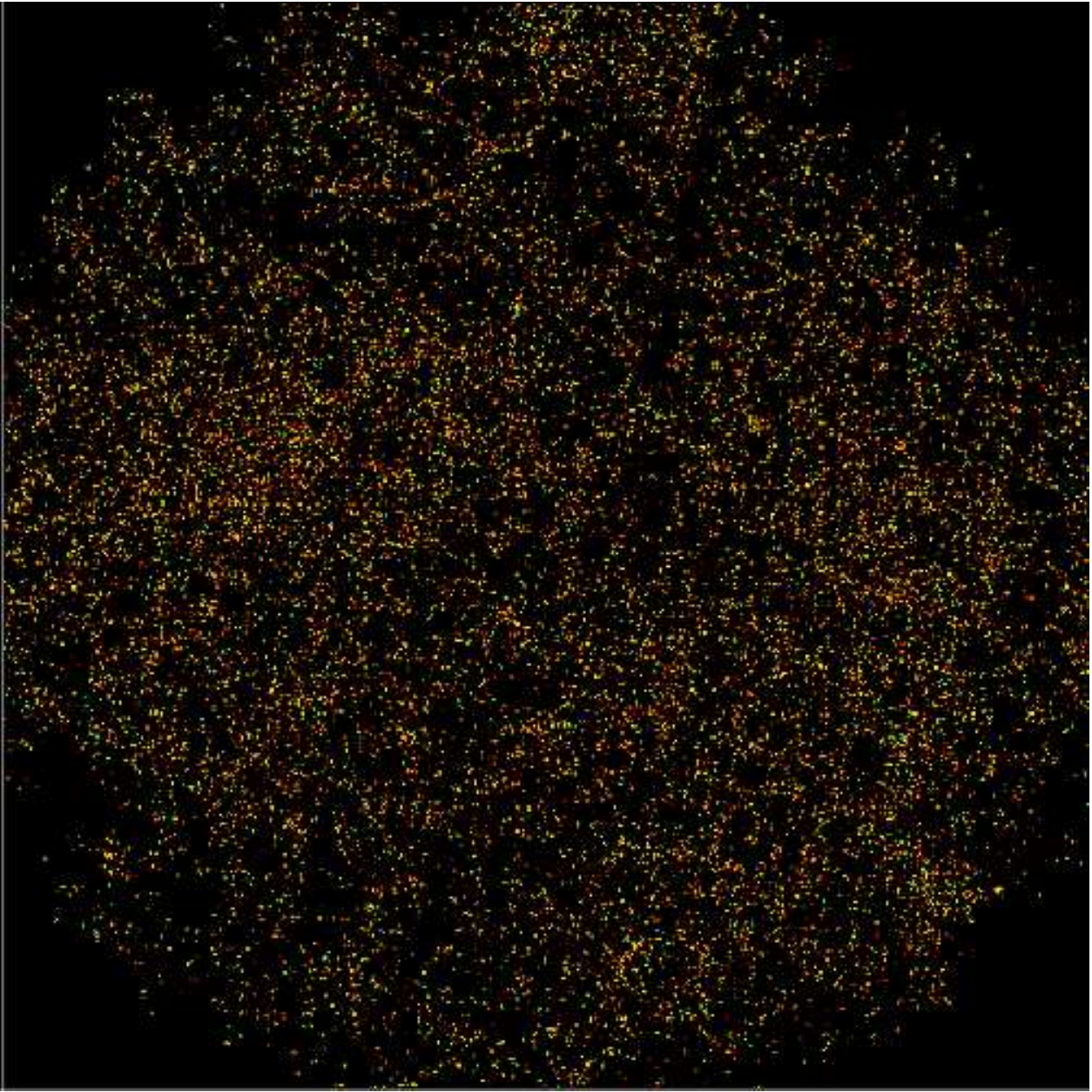}
    \end{minipage}
    \begin{minipage}[t]{0.49\textwidth}
        \includegraphics[width=1.0\textwidth]{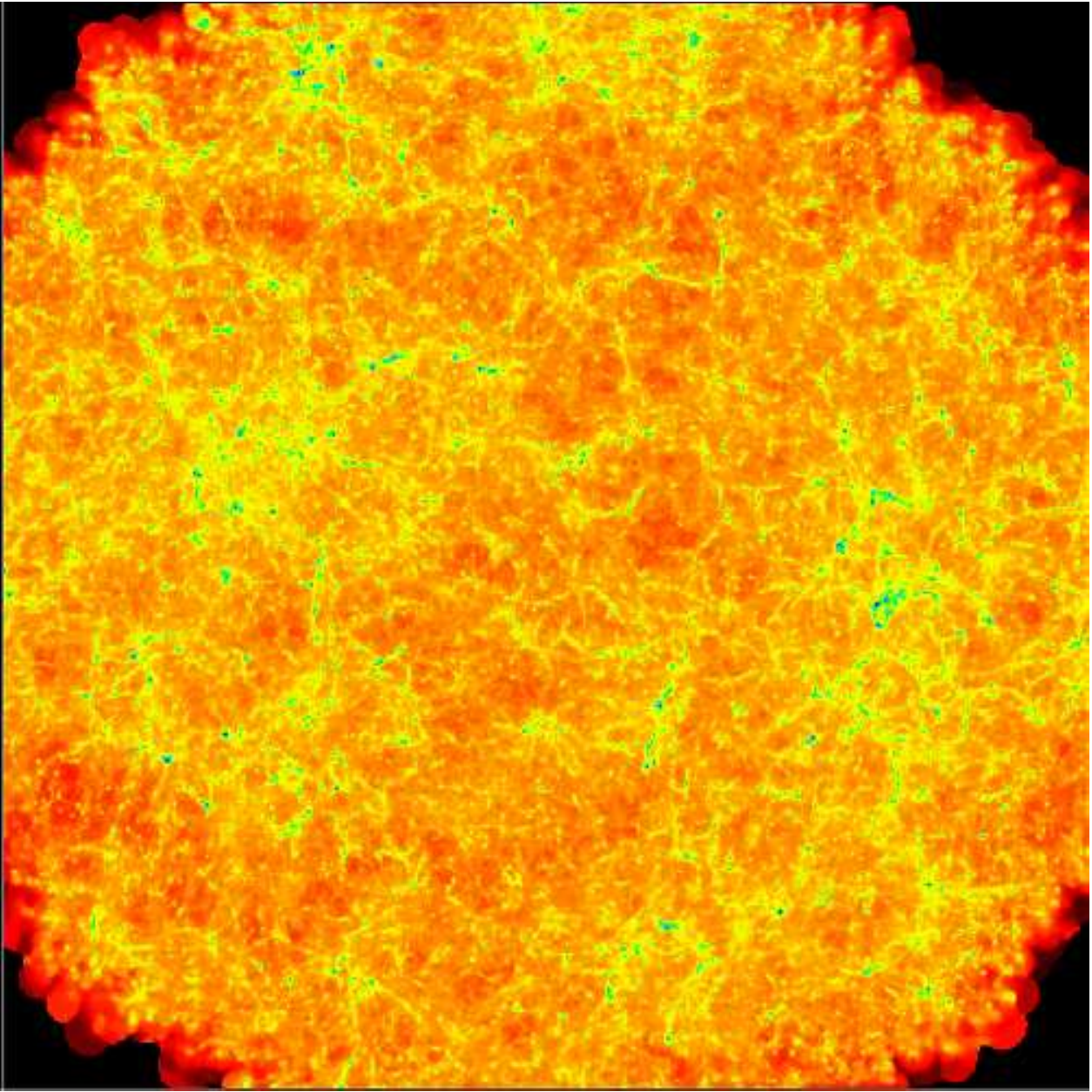}
    \end{minipage}
    \begin{minipage}[t]{0.49\textwidth}
        \includegraphics[width=1.0\textwidth]{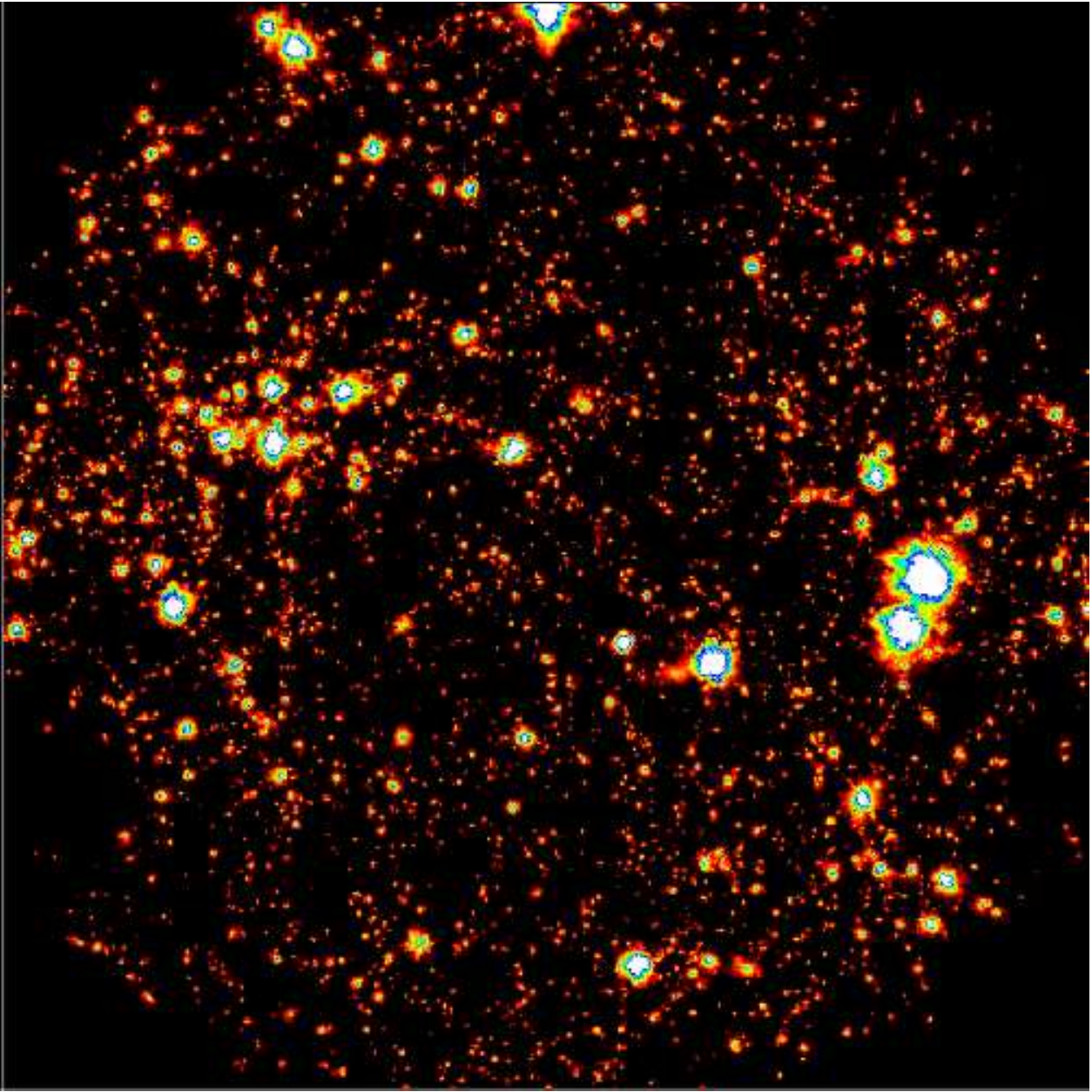}
    \end{minipage}
    \begin{minipage}[t]{0.49\textwidth}
        \includegraphics[width=1.0\textwidth]{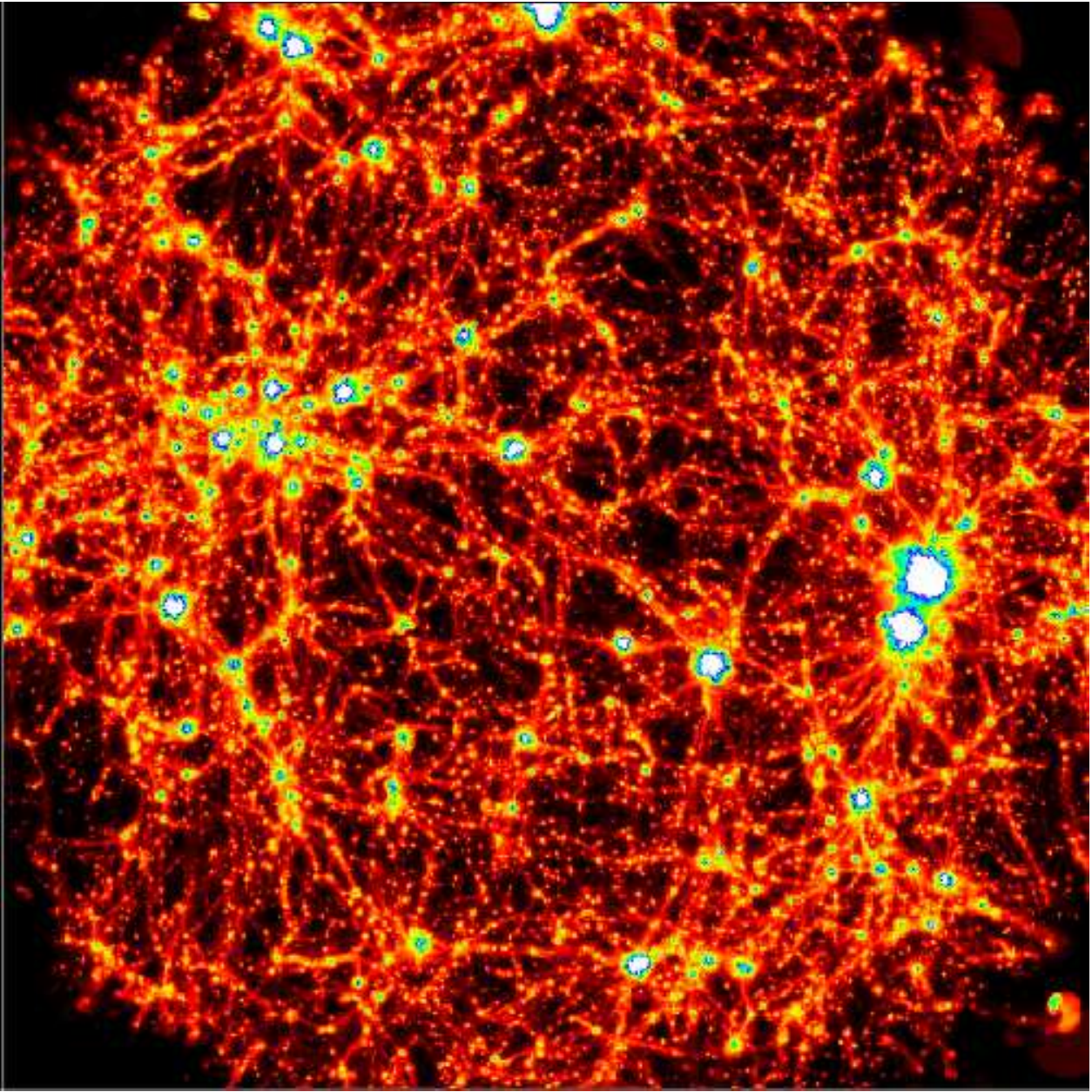}
    \end{minipage}
\caption{Visualisation of the magnetic field strength in the
  simulation box at redshifts $z=4.1$ and $z=0$. Every image shows a
  region of a linear size of $204\,\mathrm{Mpc}/(1+z)$ and was made
  using the same color bar. The upper left panel shows the magnetic
  field due to instantaneous seeding with a dipole M82-like structure
  at this redshift with maximum field strengths of $\approx 5\,
  \mathrm{nG}$. The upper right shows a simulation with homogenous
  cosmological magnetic seed field \citep{2005JCAP...01..009D}.  Here,
  the field strength more continuously traces the underlying
  structures of the matter distribution and reaches values of up to
  $\approx 10\,\mathrm{nG}$ in the highest density regions. The lower
  two panels show both simulations (left {\it Dipole}, right {\it
    Control Run}) evolved to $z=0$.}\label{vis}
\end{figure*}

\section{General results}\label{results_all}

\begin{figure*}
    \centering
    \begin{minipage}[t]{0.41\textwidth}
        \includegraphics[width=1.0\textwidth]{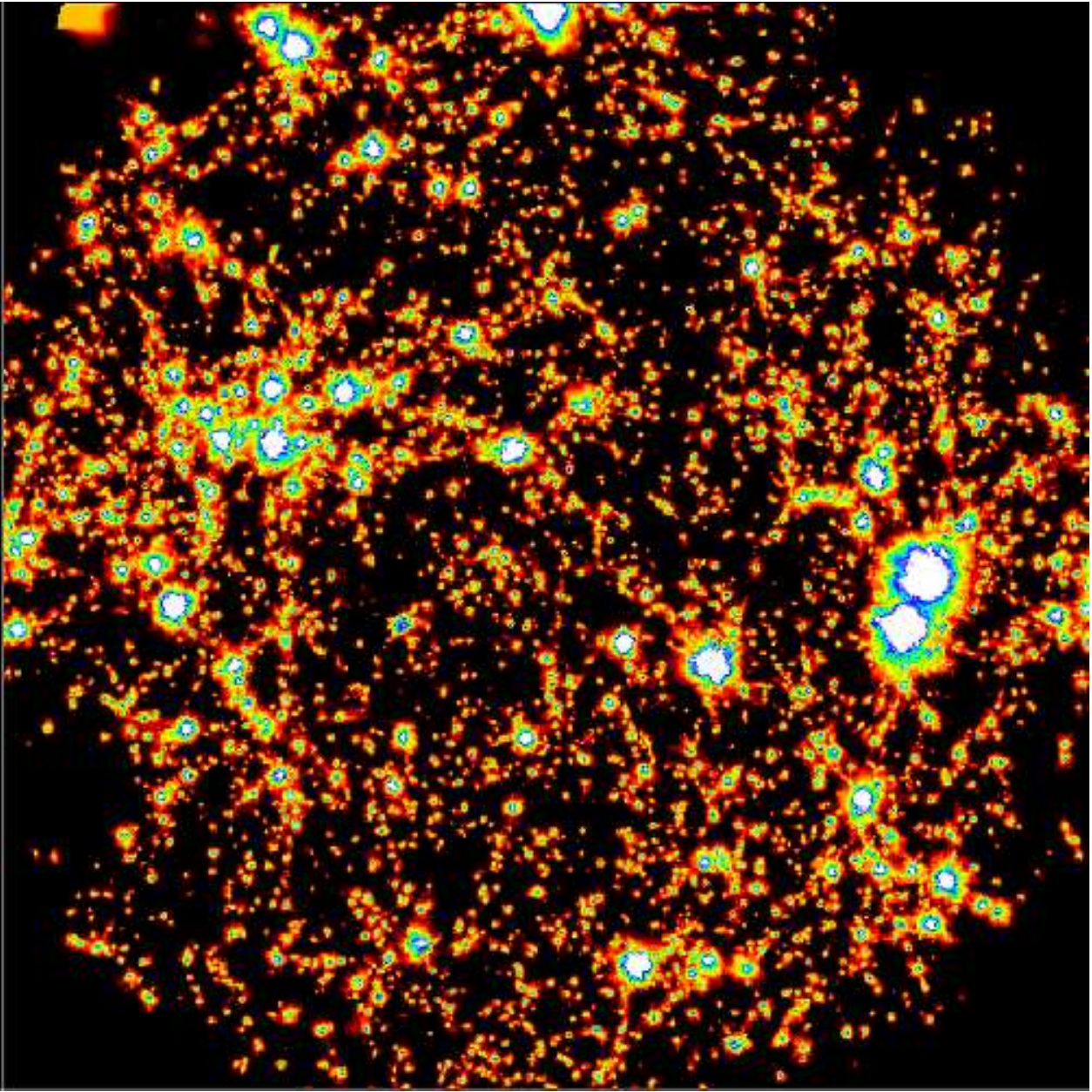}
    \end{minipage}
    \begin{minipage}[t]{0.41\textwidth}
        \includegraphics[width=1.0\textwidth]{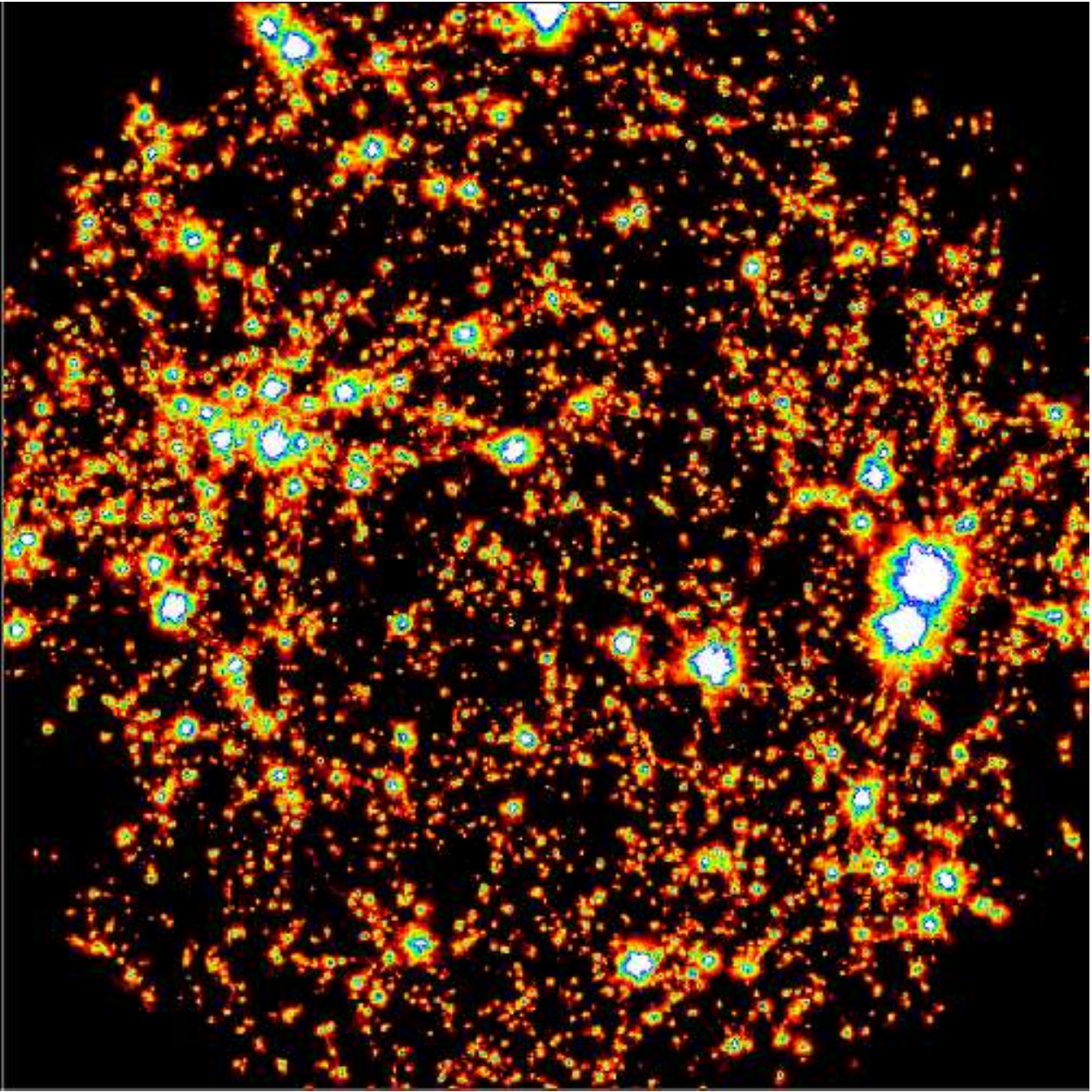}
    \end{minipage}
    \begin{minipage}[t]{0.41\textwidth}
        \includegraphics[width=1.0\textwidth]{extra/gal3_30.pdf}
    \end{minipage}
    \begin{minipage}[t]{0.41\textwidth}
        \includegraphics[width=1.0\textwidth]{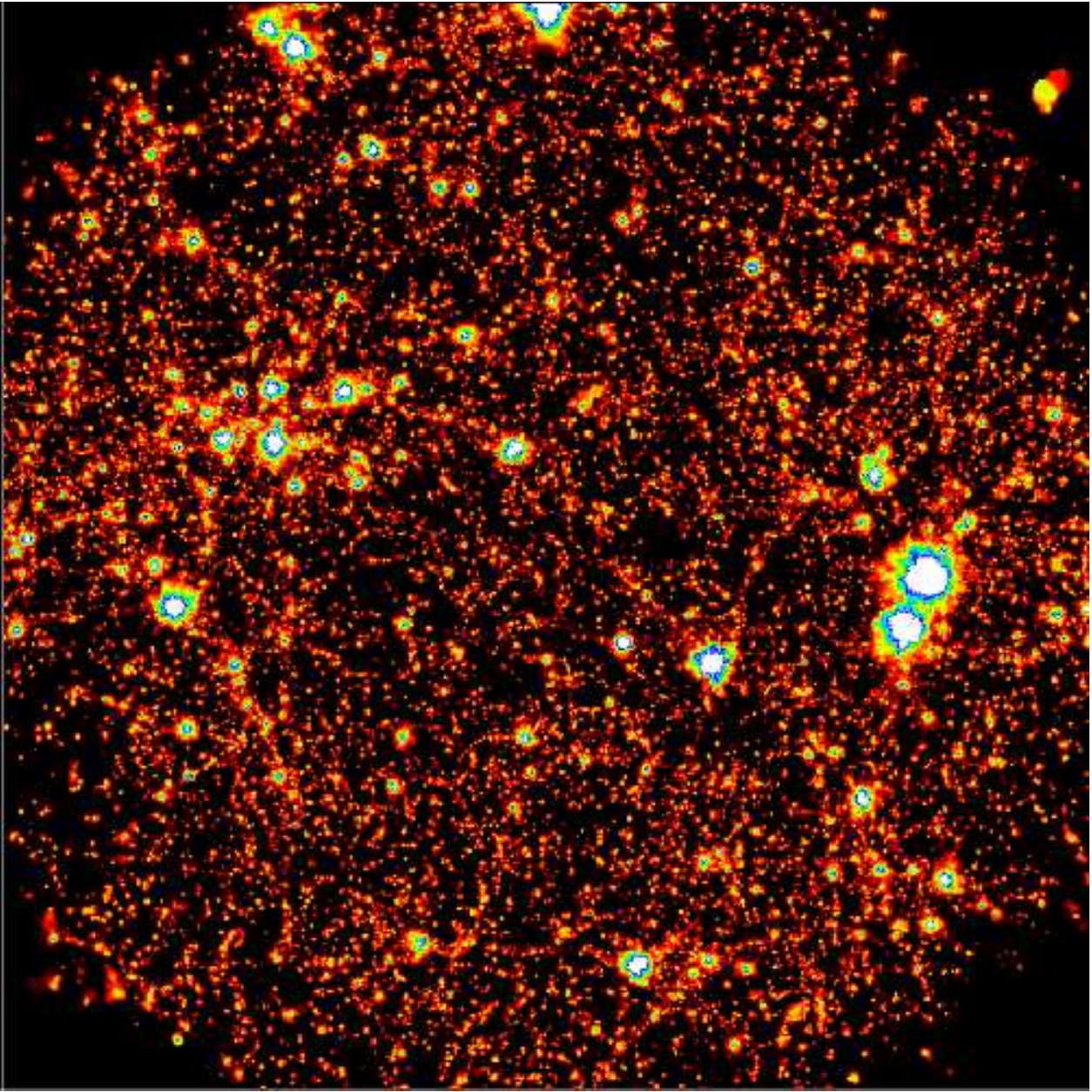}
    \end{minipage}
    \begin{minipage}[t]{0.41\textwidth}
        \includegraphics[width=1.0\textwidth]{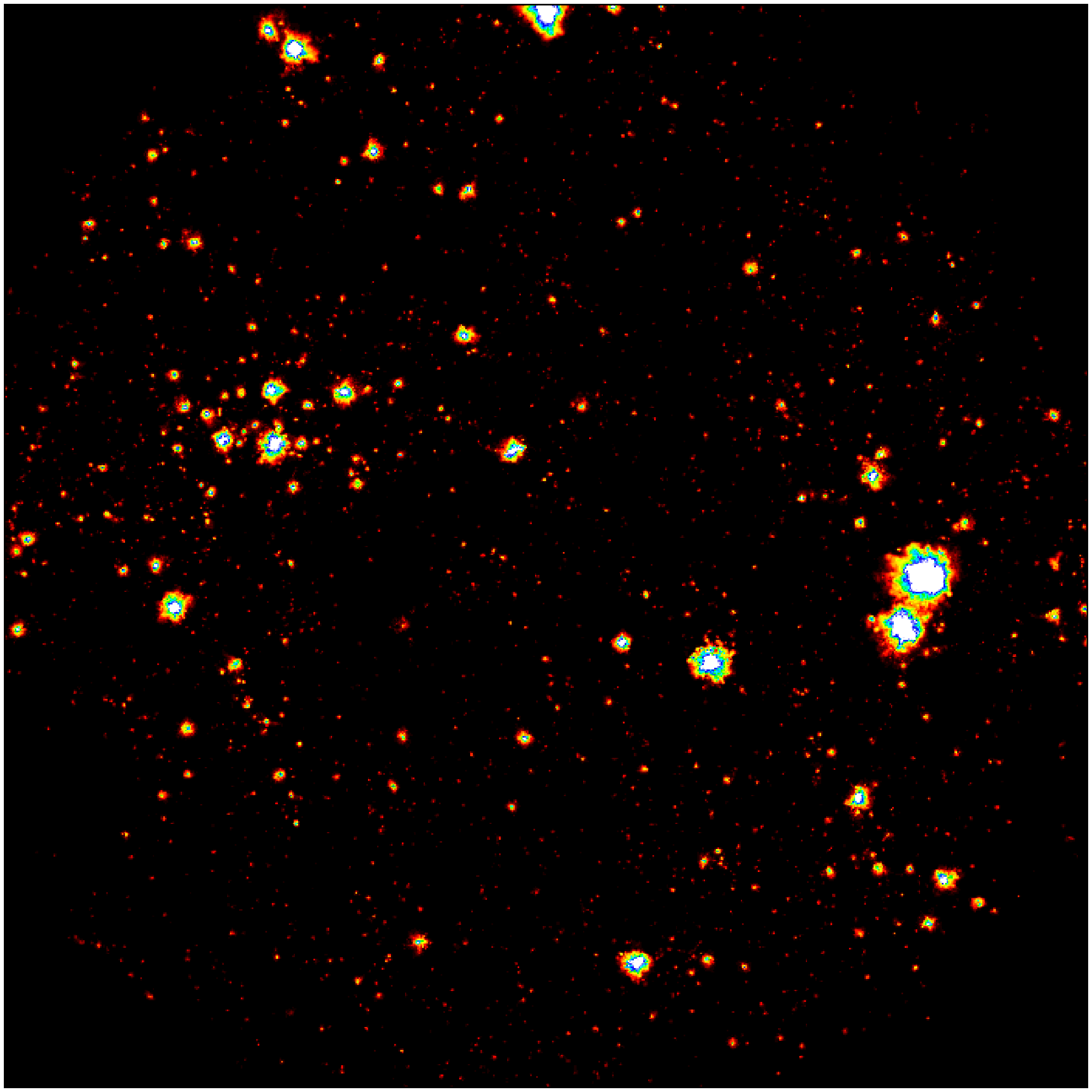}
    \end{minipage}
    \begin{minipage}[t]{0.41\textwidth}
        \includegraphics[width=1.0\textwidth]{extra/mhdz_30.pdf}
    \end{minipage}
\caption{Visualization of the magnetic field strength in the
  simulation box at redshift $z=0$. Every image shows a region of
  $204\,\mathrm{Mpc}/(1+z)$, using the same arbitrary color bar.
  Shown are the results of the {\it Dipole} (top left), {\it 0.1
    Dipole} (middle left), {\it 0.01 Dipole} (bottom left),{\it
    Quadrupole} (top right), {\it Multi Seed} (middle right), and the
  {\it Control} simulation (bottom right), respectively.}\label{vis30}
\end{figure*}

Figure \ref{vis} shows the results from the two different seeding
strategies compared in this work.  The right column shows the case of
primordial magnetic seed field ({\it Control Run}) and the left column
a run which used the seed fields of galactic outflows ({\it
  Dipole}). Shown is the magnetic field amplitude projected through
the box. The top row shows the field at the first seeding of the
galactic outflows while the lower row shows the final magnetic field
at redshift zero. One can clearly see the differences between the two
models at early times: the low-level primordial magnetic field fills
the whole volume, whereas the magnetic field from the galactic
outflows is concentrated on the galaxy population. At low redshift the
magnetic field in galaxy clusters appears comparable in both
models. This reflects the fact that inside the galaxy clusters the
magnetic field is strongly processed by compression, shear flows and
turbulence.  Therefore, the final magnetic field is shaped by such
processes rather than by the initial conditions, in agreement with
previous findings \citep{1999A&A...348..351D,2002A&A...387..383D} and
analytical modeling of saturated dynamos in clusters
\citep{2006MNRAS.366.1437S,2006A&A...453..447E}.  However, the
magnetic field in filaments looks quite different.  In the model with
galactic outflows the magnetic field in the collapsed objects along
the filaments is higher than for a primordial seed field; the magnetic
field inside the diffuse component of the filaments is much more
prominent in the case of cosmological seed fields,

 because the density in the filaments is too low to strip the hot
  and magnetized atmosphere of the galactic halos completely.  Note
  that our models do not include a kinematic component which could
  lead to an evaporation of the haloes around galaxies. This and
  especially the effect of AGN-driven outflows could change the
  picture, as the outflows could fill volumes of Mpc size in these low
  density environments (see \citet{2006AN....327..517K}, and
  references therein). We also note that in principle small galactic
  halos (down to masses of $10^8 M_\odot$) -- which are not resolved
  in the simulation -- could contribute to the magnetization of the
  low density environment. However, observations suggest that the
  star-formation rate of such galaxies would be quite small (far below
  the value of $10 M_\odot/{\mathrm yr}$ \citet{2007ApJ...670..156D}
  adopted in this work for resolved galaxies). Wence, we do expect
  them to contribute significantly.

Figure \ref{vis30} shows the magnetic field amplitude projected
through the box for all different runs at redshift zero. When the halo
magnetic field in the wind model is reduced, the field within the
individual galaxies is strongly reduced. Therefore the magnetic fields
within the filaments become less prominent ({\it Dipole}, {\it 0.1
  Dipole} and {\it 0.01 Dipole} panels).  However, the magnetic field
within the collapsed structures of galaxy clusters only slightly
changes, indicating that the amplification within the cluster centers
is strongly influenced by saturation effects, well in agreement with
previous findings \citep{2005JCAP...01..009D}. The situation changes
for the multi seed run ({\it Multi Seed}). Although again the magnetic
field in the clusters is not affected, a clear change is visible in
the filaments, where many more seed fields from galaxies appear. This
is because galaxies are formed earlier in high density environments
like proto-cluster regions than in low density environments, like
filaments. Subsequent seeding of newly formed haloes therefore
preferentially happens in low density environments like filaments.

\begin{figure}
    \centering
    \includegraphics[width=0.5\textwidth]{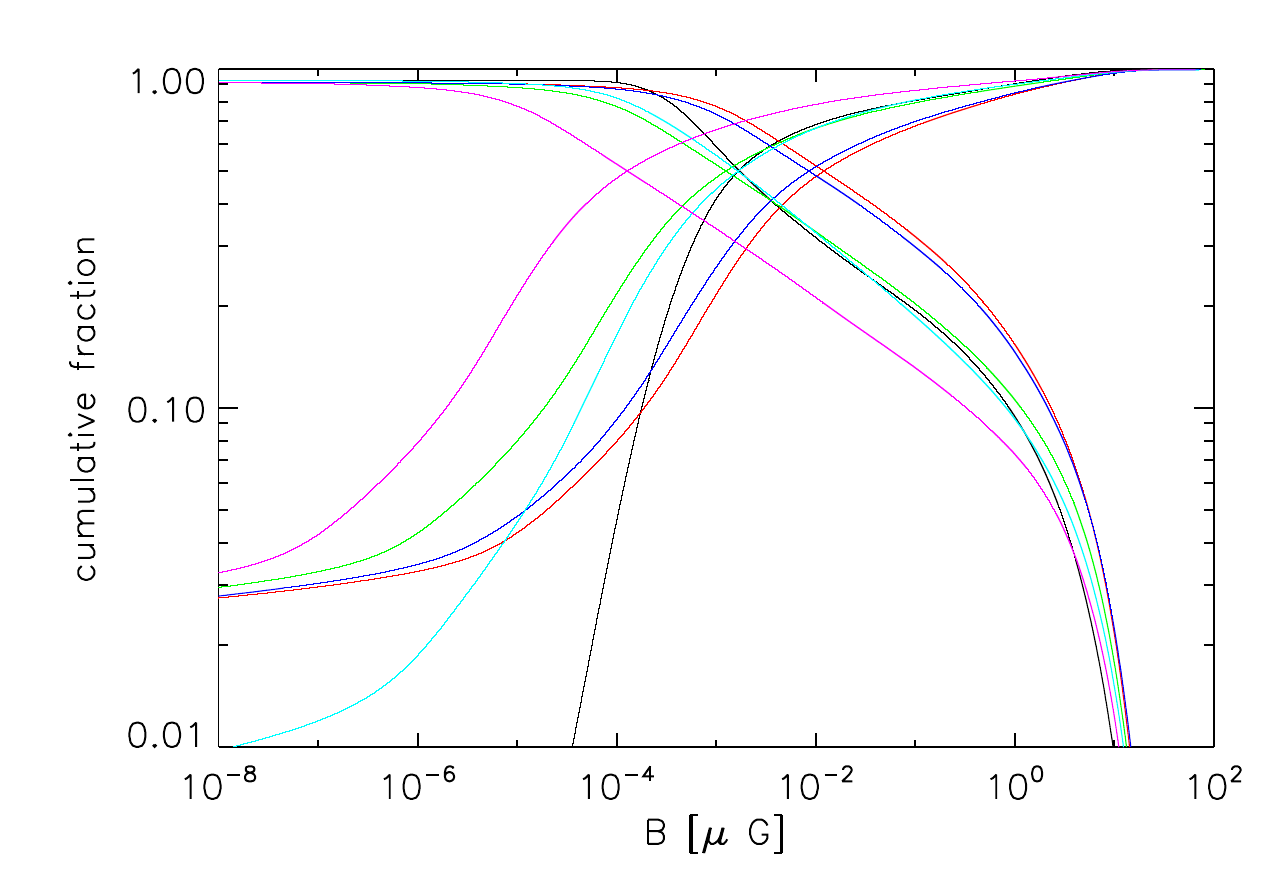}
\caption{Volume-weighted cumulative filling factor of the magnetic
  field in the cosmological box. The filling factor is plotted above
  and below a threshold as a function of the threshold (the rising and
  falling curves).  The black lines show the cosmological seed field
  from \citet{2005JCAP...01..009D}.  Also shown are the wind-induced
  seeding from the {\it Dipole} (red) and {\it Quadrupole} (blue) and
  the curves from the simulations with decreased field strength (green
  curve: {\it 0.1 Dipole}; magenta curve: {\it 0.01 Dipole}) . The
  results from the simulations with the {\it Multi Seeds} are given by
  the light blue curve. \label{fillf}}
\end{figure}

A more quantitative way of displaying the differences of our models is
to show the volume weighted cumulative filling factors obtained within
the whole box. In figure \ref{fillf} the rising (falling) curve
indicates the relative filling of the box by a magnetic field weaker
(stronger) than a certain value, for all simulations at redshift
$z=0$. In the very high field regime ($\mu \mathrm{G}$ and above) all
simulations behave similiar to each other, again indicating that
amplification within galaxy clusters is strongly influenced by
saturation effects. In the intermediate regime (most visible in the
rising curves), the simulations split in three groups, distinguished
by the chosen value for the disc magnetic field, which controls the
amount of magnetic field within the galactic outflow. In this regime,
magnetic field amplification reflects the underlying velocity field
structure. The simulation, which evolves a cosmological seed field
({\it Control Run}), corresponds (by chance) to a medium value of the
galactic fields ({\it 0.1 Dipole}).  At low magnetic field (best
visible in the rising curves), the filling factor is dominated by the
properties of the seed field.  Therefore, all models are clearly
distinguishable. This region corresponds to the low density regions,
including the diffuse matter in filaments. For lowest fields, the four
single seeded simulations converge to a non zero value, representing
the un-magnetized volume left in the box.  The multi seeded simulation
shows a lower fraction of unmagnetized volume, which is expected from
the seeding strategy.  In the simulation using a cosmological seed
field, no volume is filled with a field smaller than this seed field
(adiabatically diluted with the local density).

This shows that for galaxy clusters the single seed simplification in
our approach is not crucial. On the other hand, low density regions
like filaments provide a natural testbed for the origin of
cosmological magnetic fields and also are very sensitive to the
approximations and simplifications made.

 This is expected from previous findings. In simulations where a
  cosmological magnetic field is evolved, the field amplification is
  linear in low density regions undergoing mainly adiabatic
  compression or dilution, whereas the orientation is typically
  aligned with the filament
  (\citet{2005JCAP...01..009D,2005ApJ...631L..21B}).  Therefore, such
  regions are only mildly influencing the magnetic field configuration
  imposed by the seeding mechanism.

\section{Cluster Based Analysis} \label{results}

The amplification of magnetic fields in galaxy clusters is driven
by adiabatic compression, and  merger events, which induce
shear flows and subsequent turbulent flows. All these processes
depend on the cluster environment and therefore may depend on cluster mass. 
Observationally there is yet not enough data available to track how the 
magnetic field varies  within clusters as function of radius or how the 
mean field within the cluster varies with cluster mass (or equivalent, with 
the mean cluster temperature). \citet{2007MNRAS.378.1565C} concluded 
from the properties of observed, diffuse radio emission (so-called radio halos) 
that the mean magnetic field within the radio emitting region appeared 
independent of cluster mass. 

But the volume of the radio-emitting region is found to
increase with cluster mass.

 Therefore, even if the central magnetic field does increase with
  cluster mass, a constant mean magnetic field could still be
  inferred, if the shape of the radial field profile decreases
  properly.  Furthermore, although indications for a radial dependence
  of the magnetic field can be inferred from clusters where rotation
  measures of several radio galaxies are available
  \citep{2001A&A...378..777D,2004A&A...424..429M,2006AN....327..539G},
  there exits still a large degeneracy between the parameters
  describing the magnetic field in clusters
  \citep{bonafede08,2008A&A...483..699G}.  Hence, it is quite
  important to compare observed scalings with predictions from
  simulations. Especially the dependence of scaling relations on the
  adopted model for the origin of the magnetic seed fields has to be
  studied in more detail.  To this end, various simulations performed
  in the past gave quite different results.

Models based on shock induced field generation predict magnetic field
 strengths with a very weak dependence on the distance to the
cluster center and also predict a scaling of the mean magnetic field
$B \propto \sqrt{T}$ \citep{2001ApJ...562..233M}.

 In contrast, previous SPH simulations following a primordial
  magnetic seed field predict a steep radial dependence for the mean
  magnetic field, mainly following the decline of the density
  \citep{2001A&A...369...36D,2005JCAP...01..009D}.  Such steep
  profiles were confirmed with AMR simulations
  \citep{2005ApJ...631L..21B,2008A&A...482L..13D}.  SPH simulations
  also predict a very steep correlation of the mean magnetic field
  with the cluster temperature, e.g. $B \propto T^{2}$
  \citep{1999A&A...348..351D}.

\subsection{Radial profiles}

\begin{figure*}
    \centering
    \begin{minipage}[t]{0.4\textwidth}
        \includegraphics[width=1.0\textwidth]{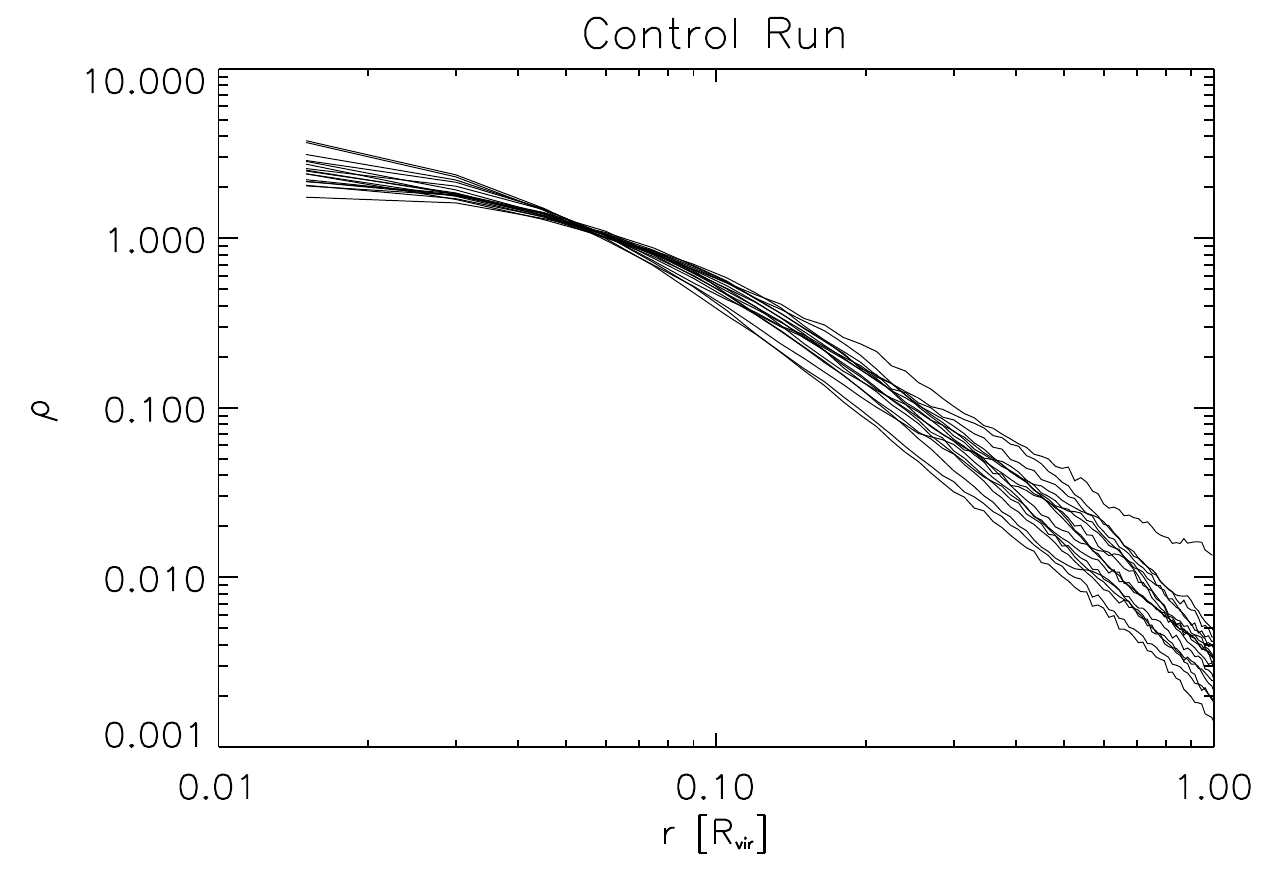}
    \end{minipage}
    \begin{minipage}[t]{0.4\textwidth}
        \includegraphics[width=1.0\textwidth]{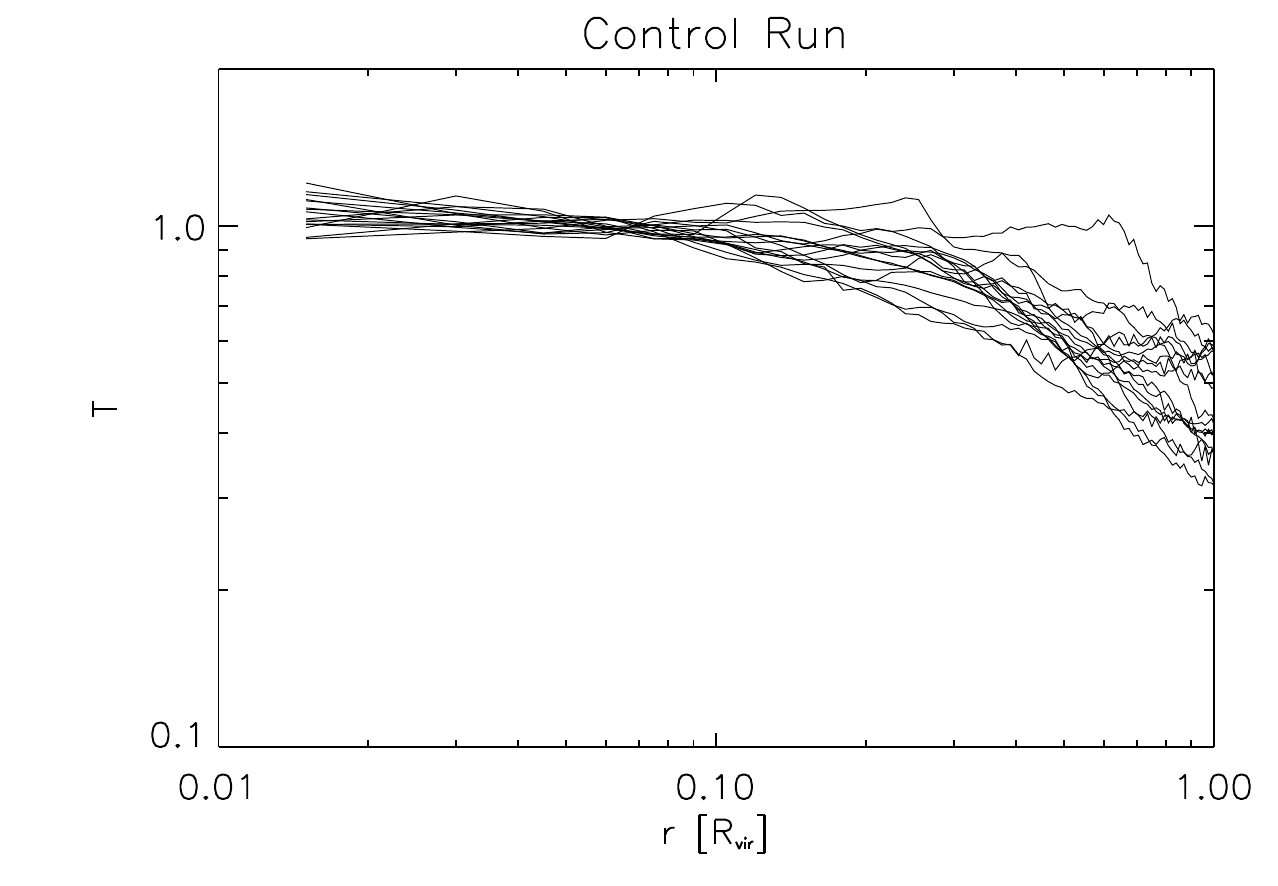}
    \end{minipage}
    \begin{minipage}[t]{0.4\textwidth}
        \includegraphics[width=1.0\textwidth]{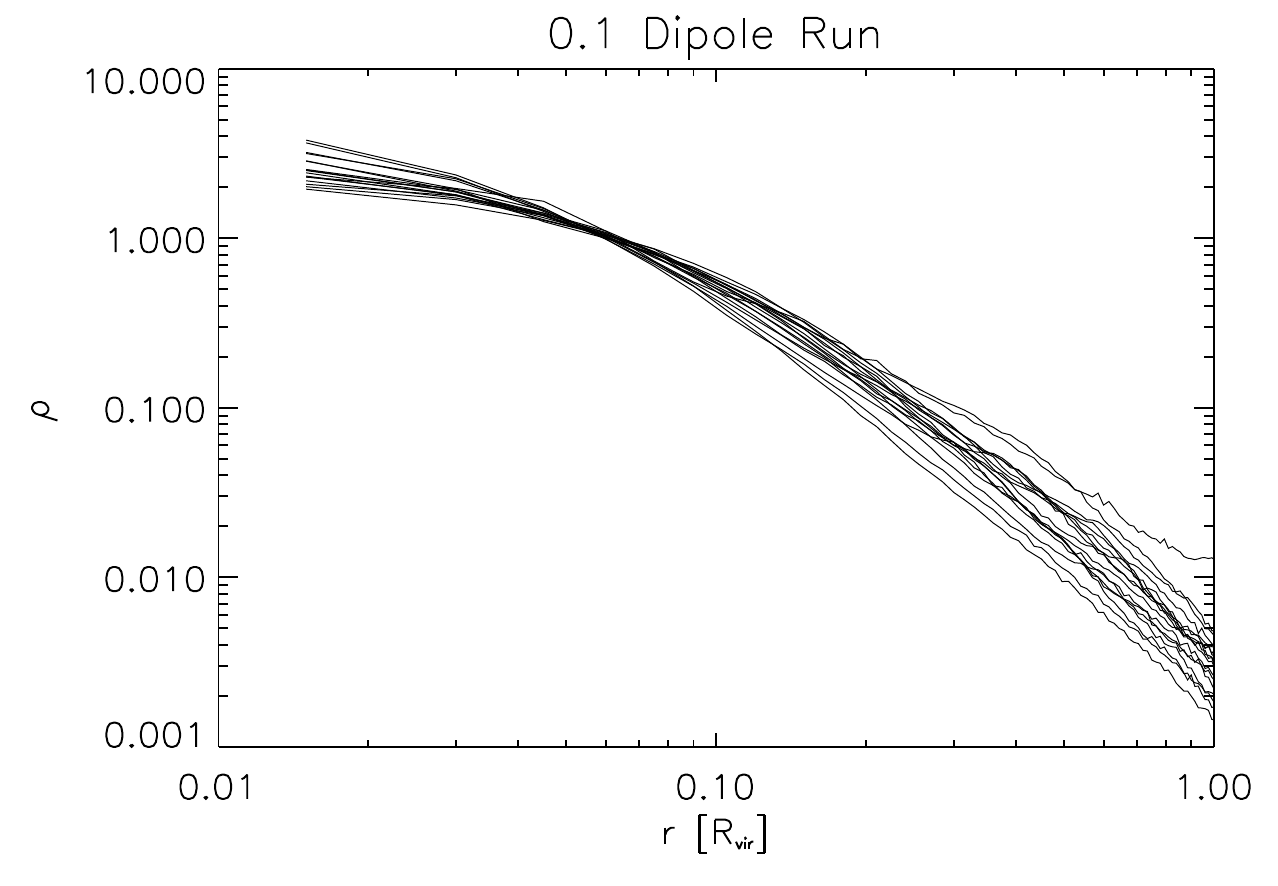}
    \end{minipage}
    \begin{minipage}[t]{0.4\textwidth}
        \includegraphics[width=1.0\textwidth]{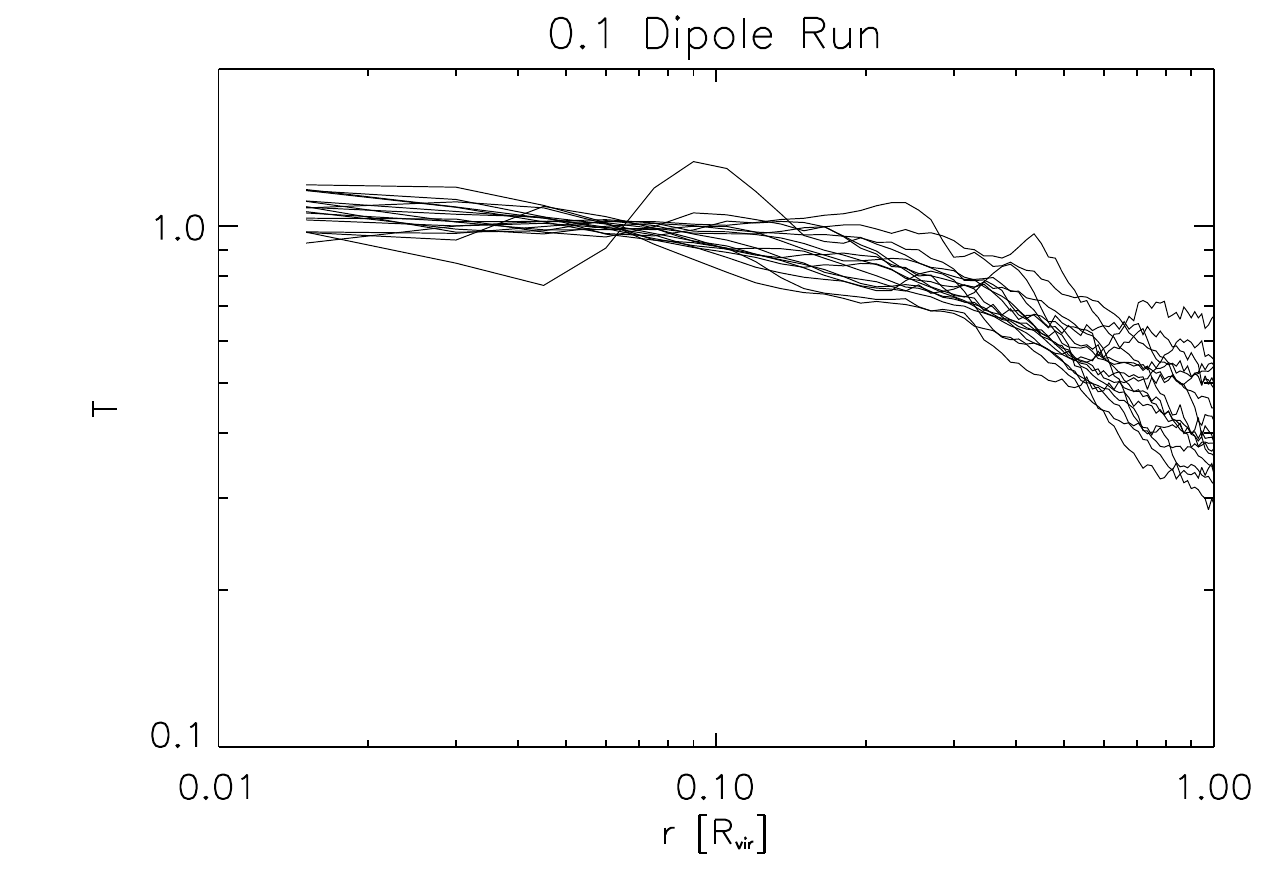}
    \end{minipage}
\caption{Density (left column) and mass weighted temperature
(right column) profiles for the 16 most massive clusters in the
simulations. Here, only the {\it Control Run} and the {\it Dipole
Run} are shown, demonstrating the absence of a significant dynamic
influence of the magnetic field in the models.}\label{rho_T_prof}
\end{figure*}

In figure \ref{rho_T_prof} we plot angularly averaged radial profiles
of density (left column) and mass-weighted temperature (right
column). The individual lines are taken from the 16 most massive
clusters within the simulation. They are scaled to the virial radius
of the individual clusters and are normalized to the same mean value
within $0.1R_\mathrm{vir}$. The upper row shows the results from the
control run following a primordial magnetic seed field
\citep{2005JCAP...01..009D}, the lower row shows results from the
dipole run with the largest value for the galactic halo
field. Although individual haloes show small variations because of the
slight changes magnetic fields introduce in the systems dynamics (by
adding small perturbations of the force field at high redshift), there
is no significant dynamical influence of the magnetic field. This is
in good agreement with the relatively small values of the magnetic
fields ($< 10\mu$G) within the simulated galaxy clusters, and also
with previous results.

\begin{figure*}
    \centering
    \begin{minipage}[t]{0.45\textwidth}
    \includegraphics[width=1.0\textwidth]{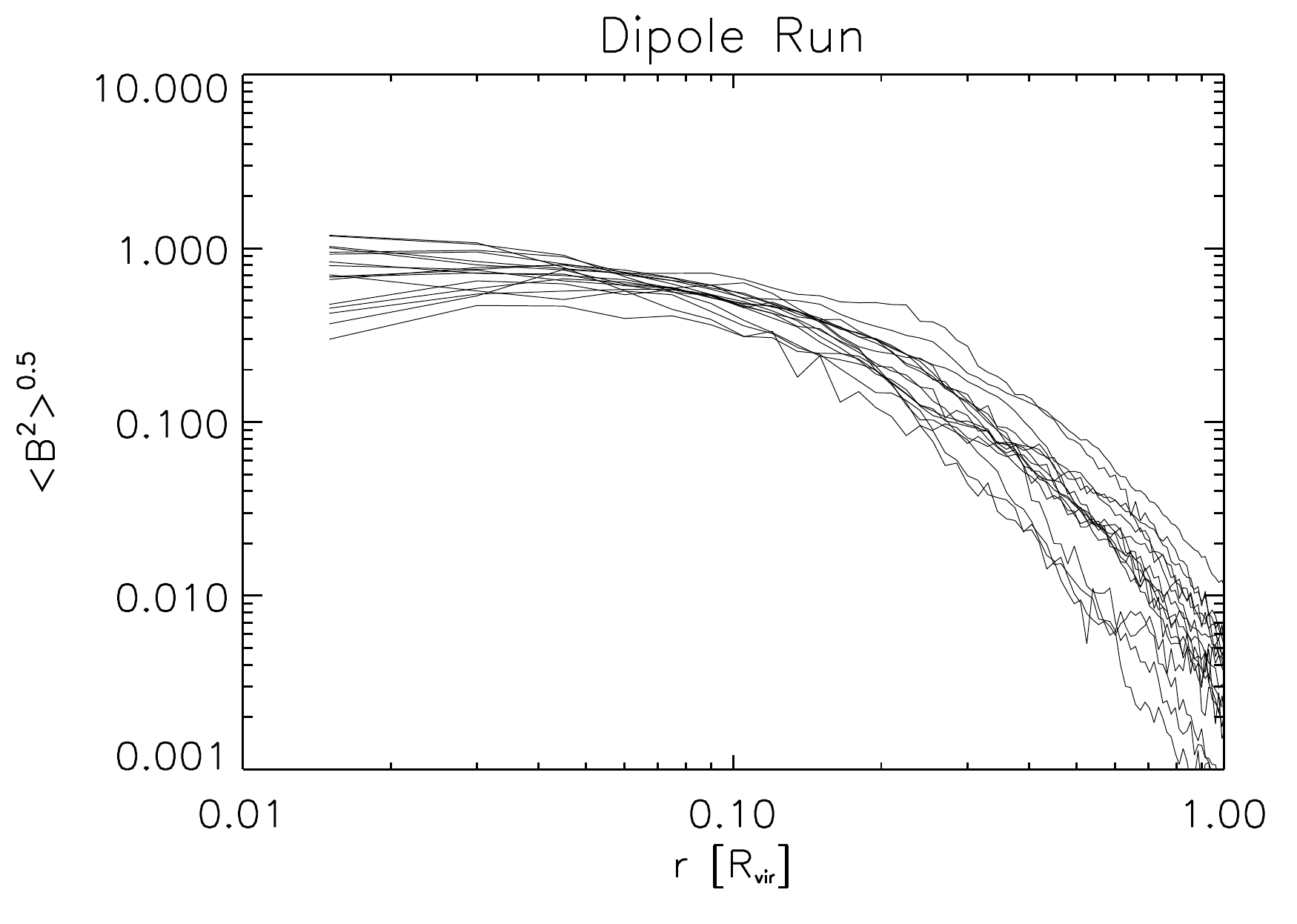}
    \end{minipage}
    \begin{minipage}[t]{0.45\textwidth}
    \includegraphics[width=1.0\textwidth]{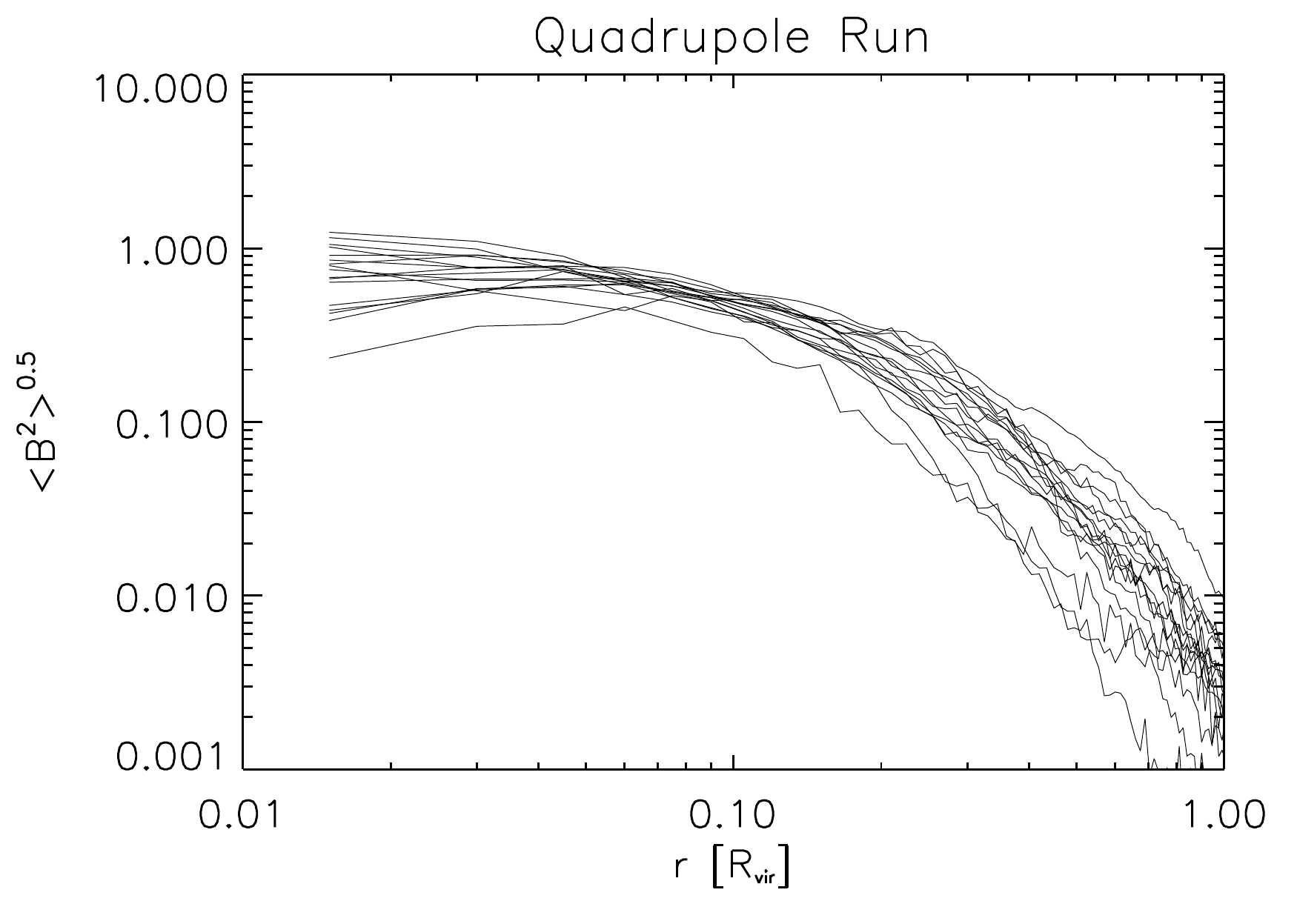}
    \end{minipage}
    \begin{minipage}[t]{0.45\textwidth}
    \includegraphics[width=1.0\textwidth]{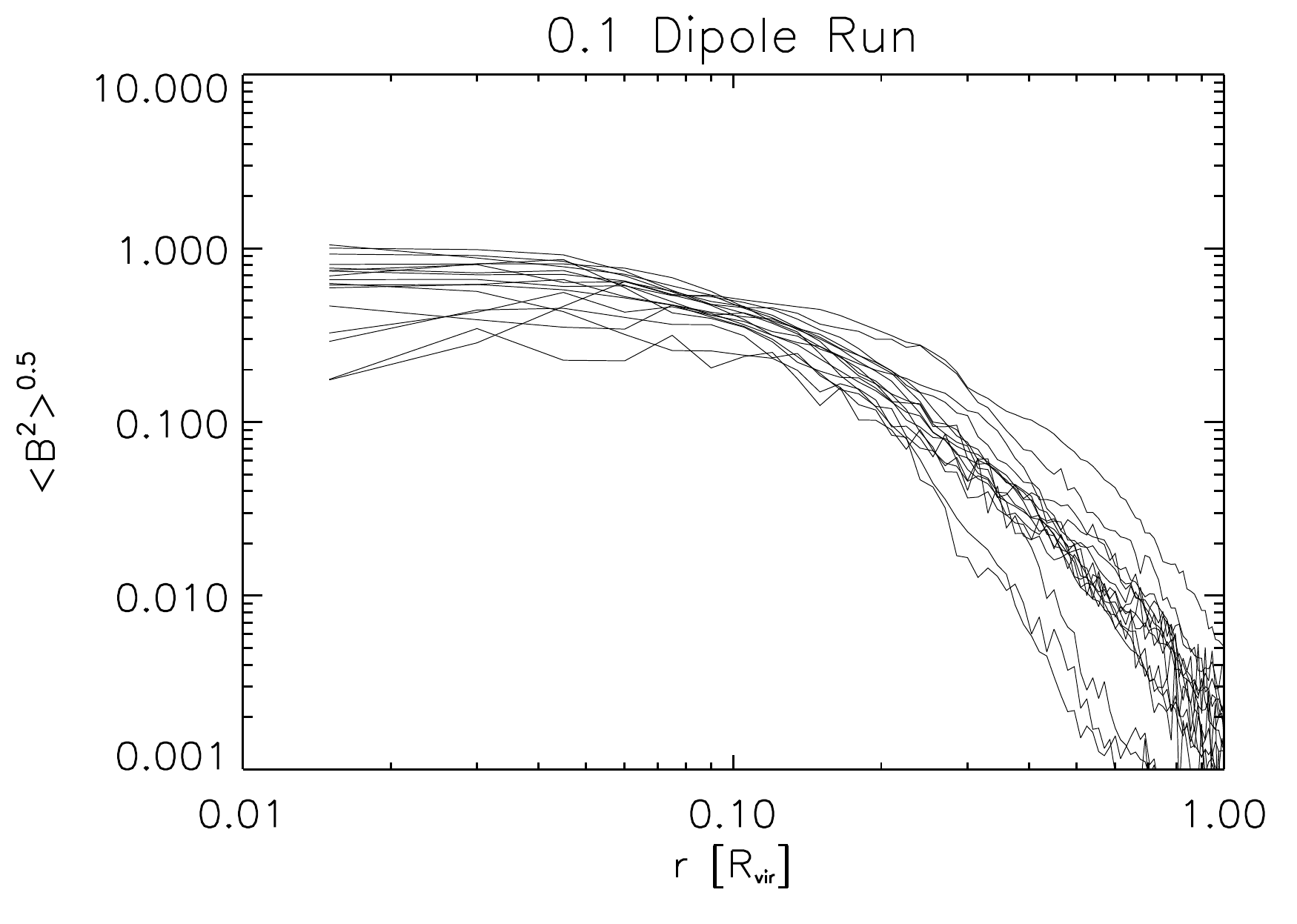}
    \end{minipage}
    \begin{minipage}[t]{0.45\textwidth}
    \includegraphics[width=1.0\textwidth]{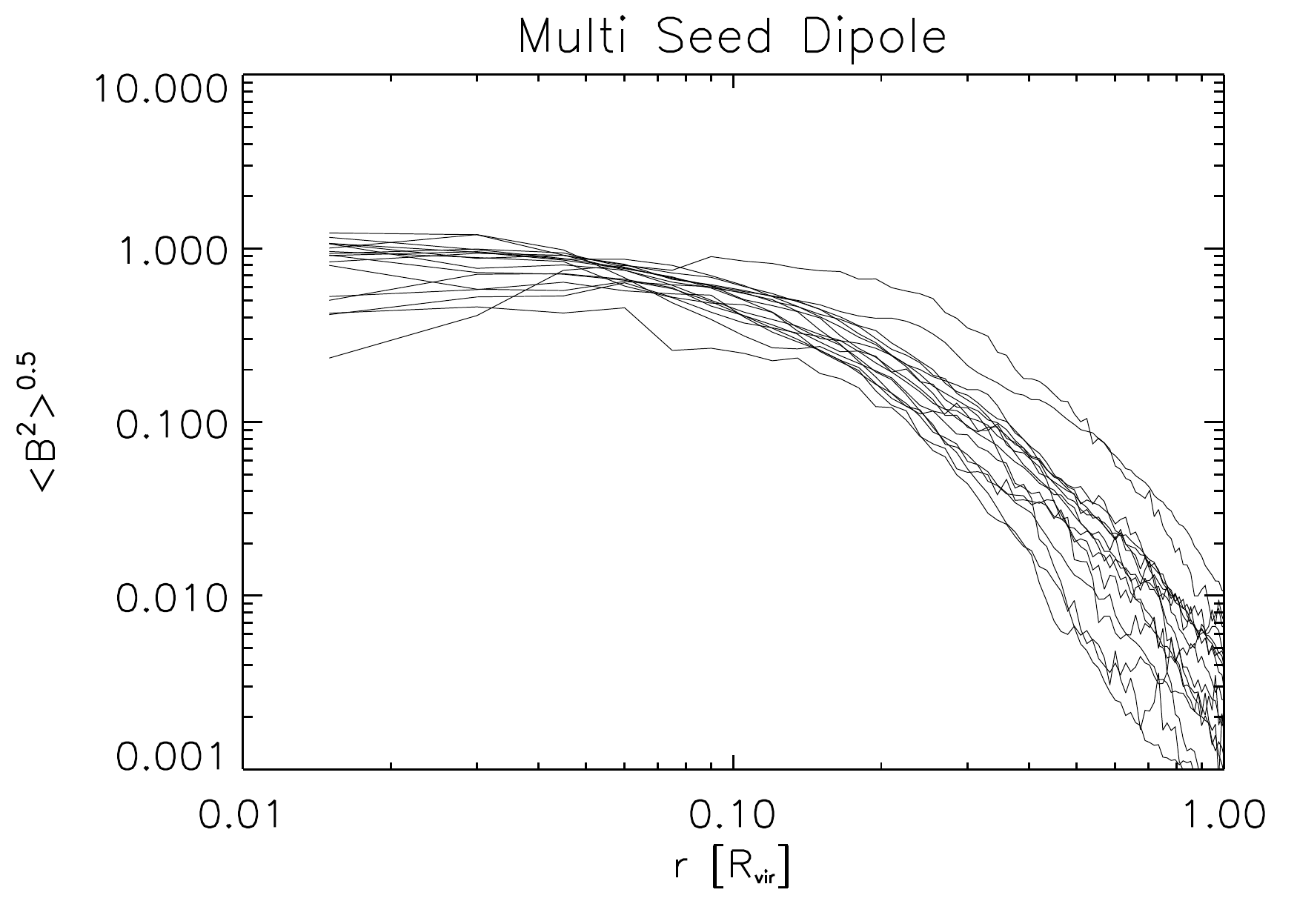}
    \end{minipage}
    \begin{minipage}[t]{0.45\textwidth}
    \includegraphics[width=1.0\textwidth]{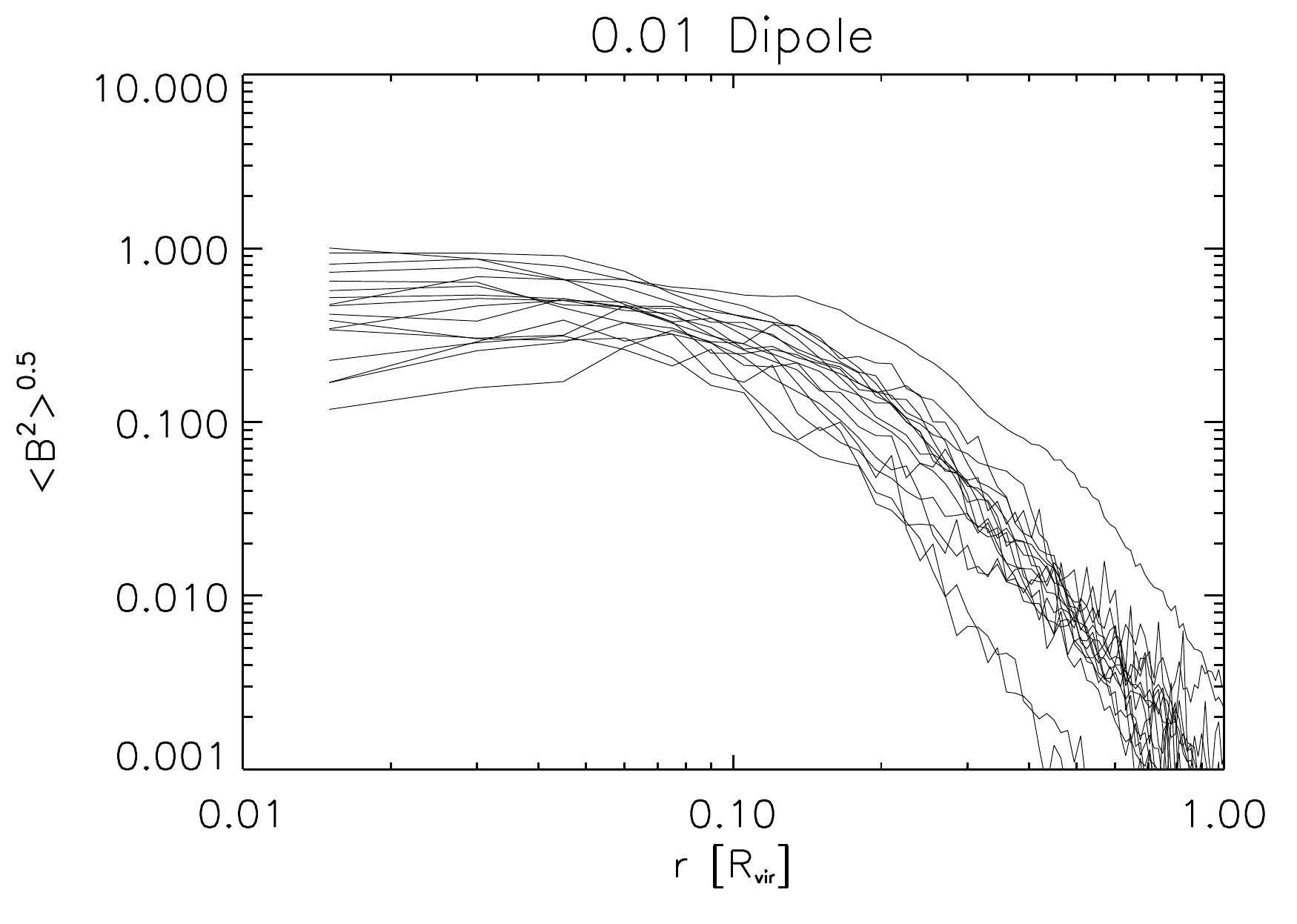}
    \end{minipage}
    \begin{minipage}[t]{0.45\textwidth}
    \includegraphics[width=1.0\textwidth]{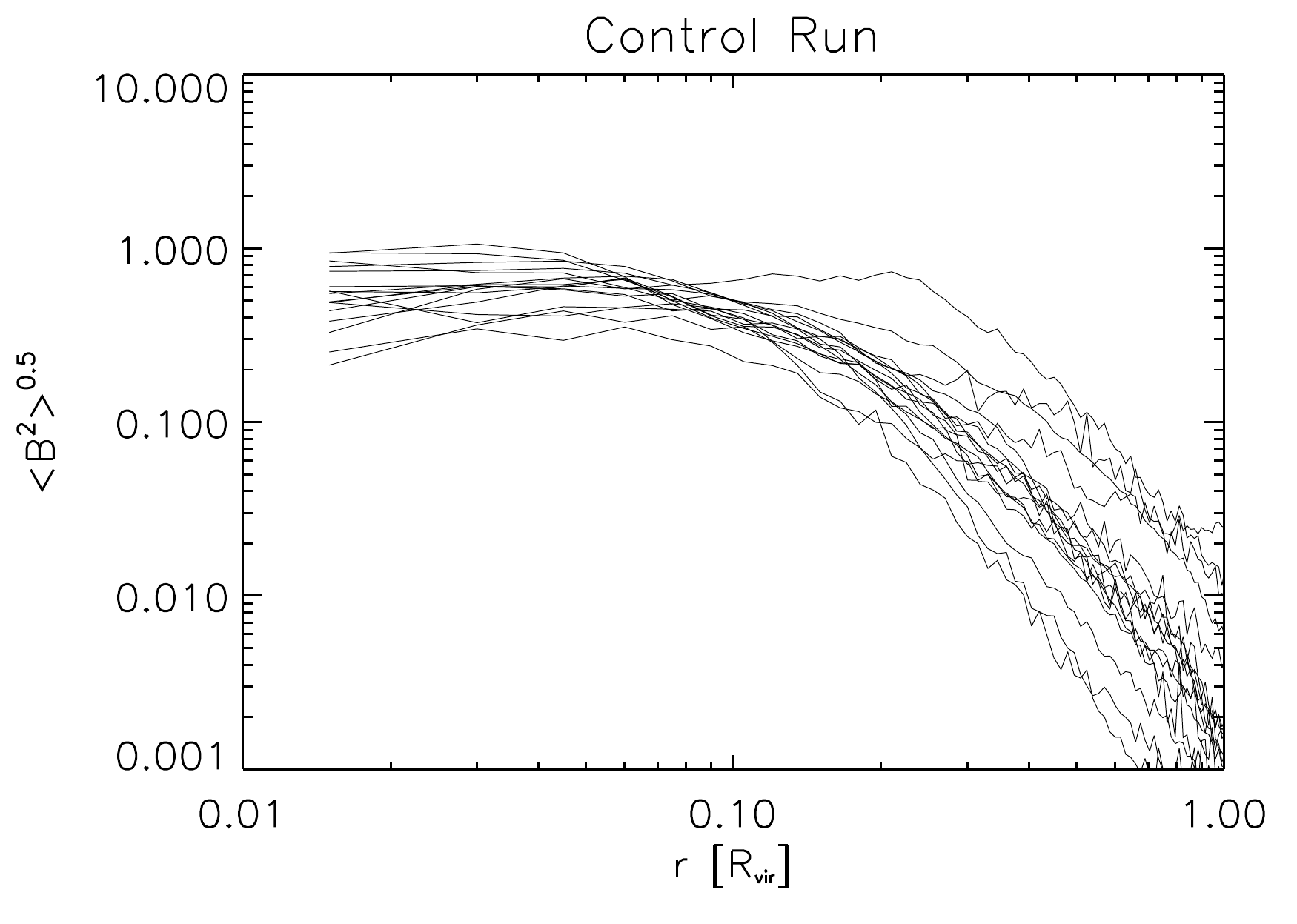}
    \end{minipage}
\caption{Radial profiles of the RMS magnetic field strength for the
16 most massive clusters. The profiles are scaled to
$R_{\mathrm{vir}}$ and normalized to the same mean value within
$0.1\times R_{\mathrm{vir}}$. Shown are the results of the {\it
Control} simulation (bottom right), {\it Dipole}
(top left),{\it 0.1 Dipole} (middleleft), {\it 0.01 Dipole} (bottom left),{\it Quadrupole} (top right) and {\it Multi Seed} (middle right), respectively.}
\label{b_prof}
\end{figure*}

Figure \ref{b_prof} shows the RMS magnetic field profiles for the
16 most massive clusters for all our simulations. As
before, they are scaled to the virial radius of the individual
clusters and are normalized to the same mean value within
$0.1R_\mathrm{vir}$. In good agreement with previous
findings, the magnetic field profiles generally follow a very similar
decline as the density, but with somewhat more scatter. However,
the scatter of the individual magnetic field profiles seems to
depend on the overall strength of the magnetic field. The
simulations with higher magnetic seed fields (e.g. the {\it
Dipole} and {\it Quadrupole} runs) show significantly less scatter.
This is a clear indication that in these simulations saturation
effects dominate the amplification of the
magnetic field in outer parts. Therefore, the total field amplification
is less dependent on the system's dynamics and so the
differences in the scaled profiles are reduced. There is also some
indication that the scatter in the profiles is slightly increased
for the {\it multiple seed} run, indicating that there could be
still some influence of the late forming galaxies on the outer
parts of the magnetic profiles. In general, aside from the scatter in
the profiles the exact details of the origin of the magnetic seed
fields do no strongly  influence the predicted shape of the radial 
profiles. Therefore, the shape of the magnetic profiles is a robust 
prediction of the simulations, driven mainly by the
dynamics imprinted from the structure formation process.

\subsection{Average field strength}

\begin{figure*}
    \centering
    \begin{minipage}[t]{0.45\textwidth}
    \includegraphics[width=1.0\textwidth]{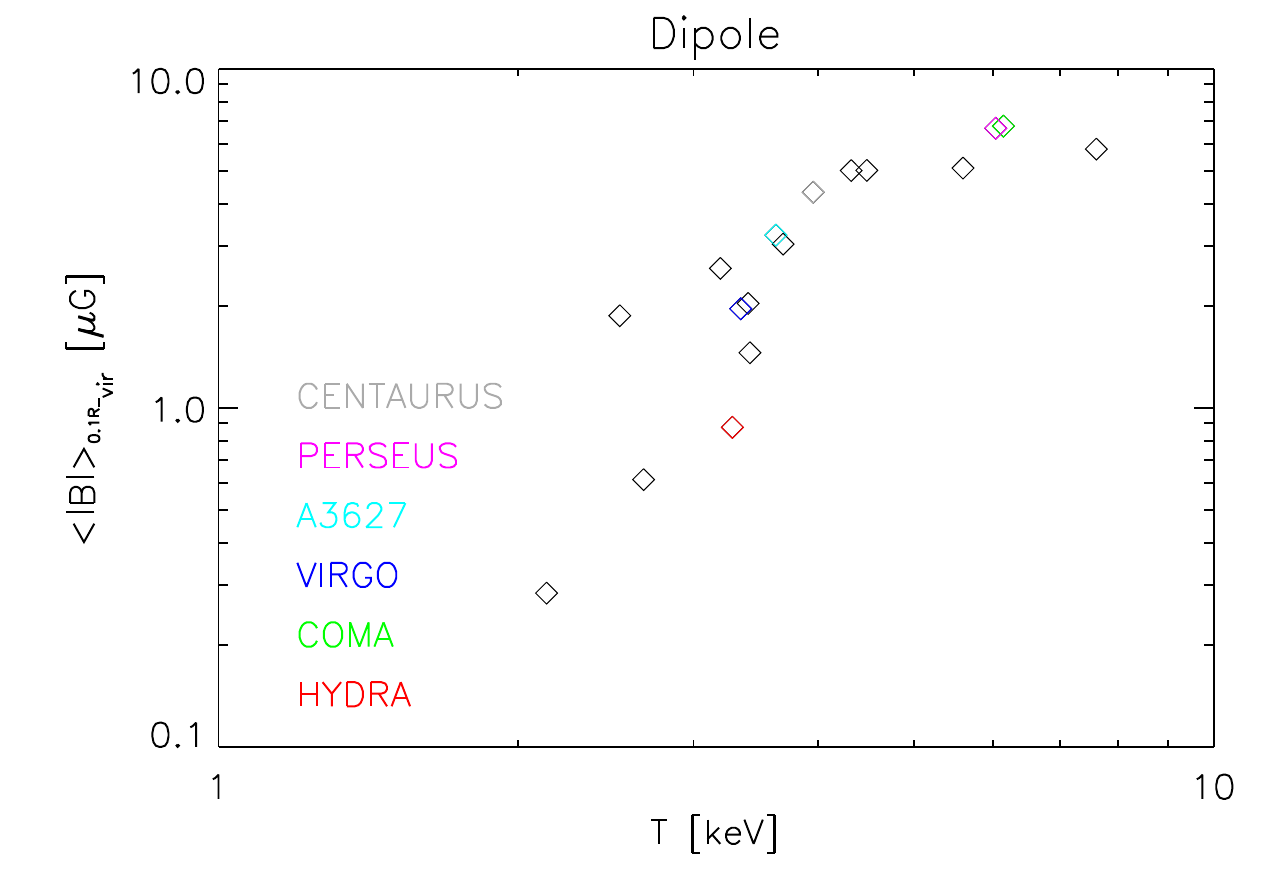}
    \end{minipage}
    \begin{minipage}[t]{0.45\textwidth}
    \includegraphics[width=1.0\textwidth]{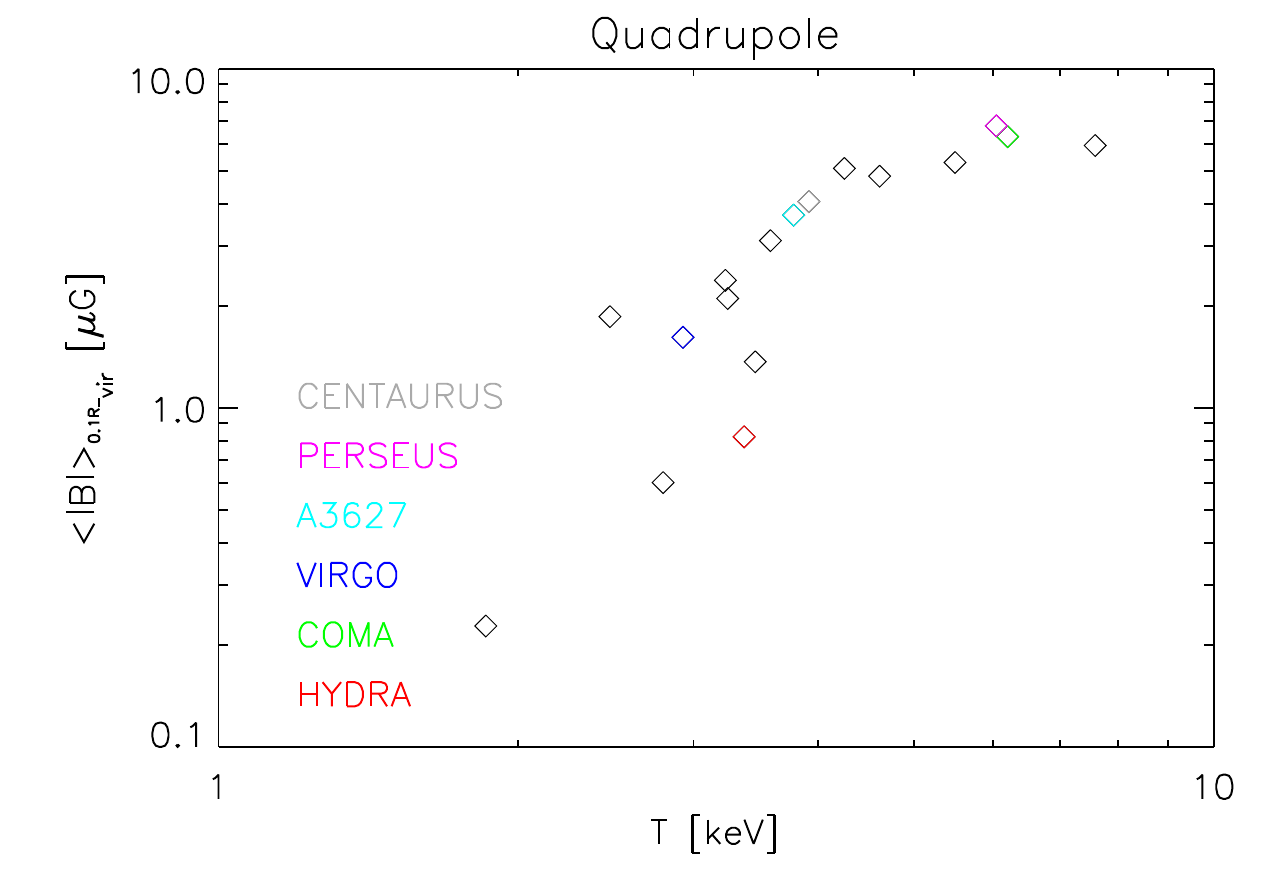}
    \end{minipage}
    \begin{minipage}[t]{0.45\textwidth}
    \includegraphics[width=1.0\textwidth]{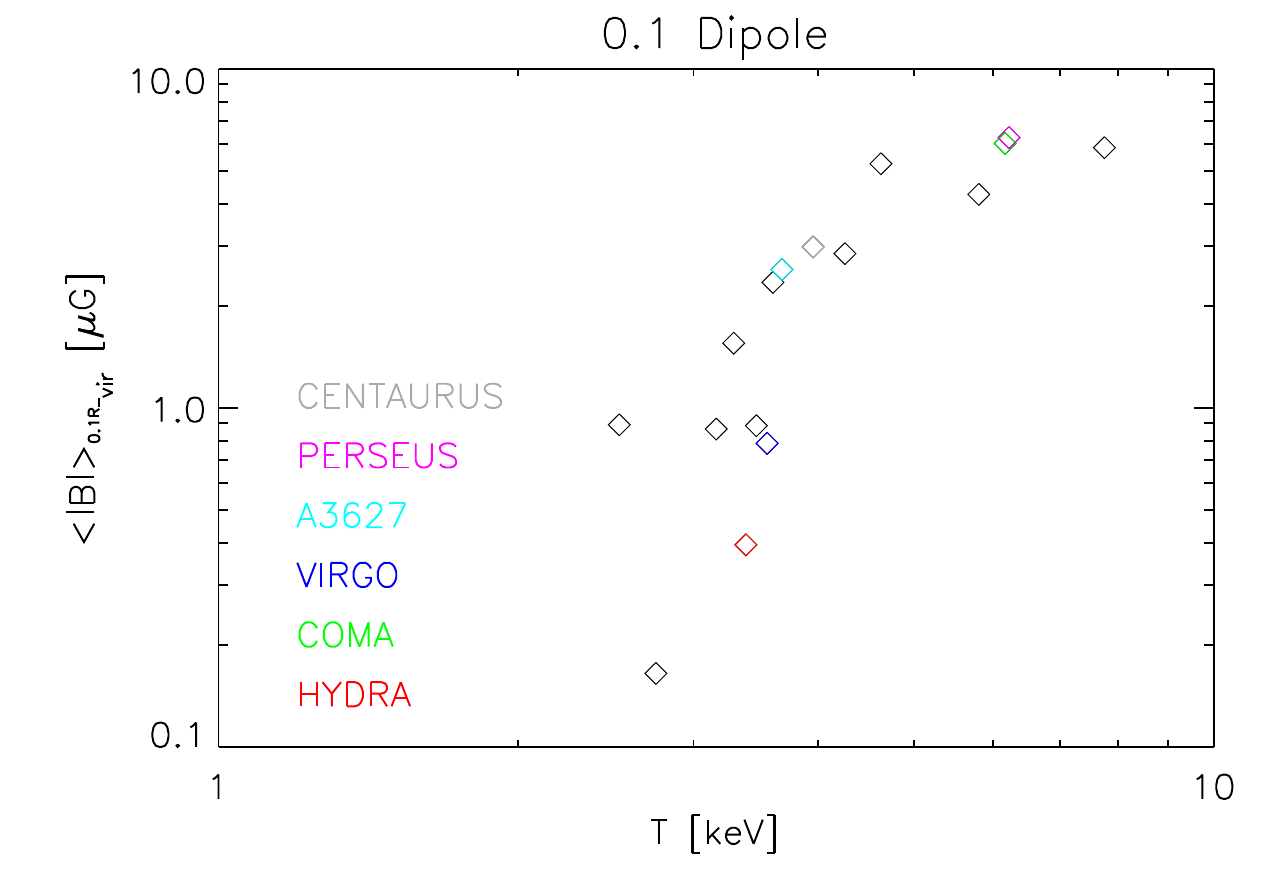}
    \end{minipage}
    \begin{minipage}[t]{0.45\textwidth}
    \includegraphics[width=1.0\textwidth]{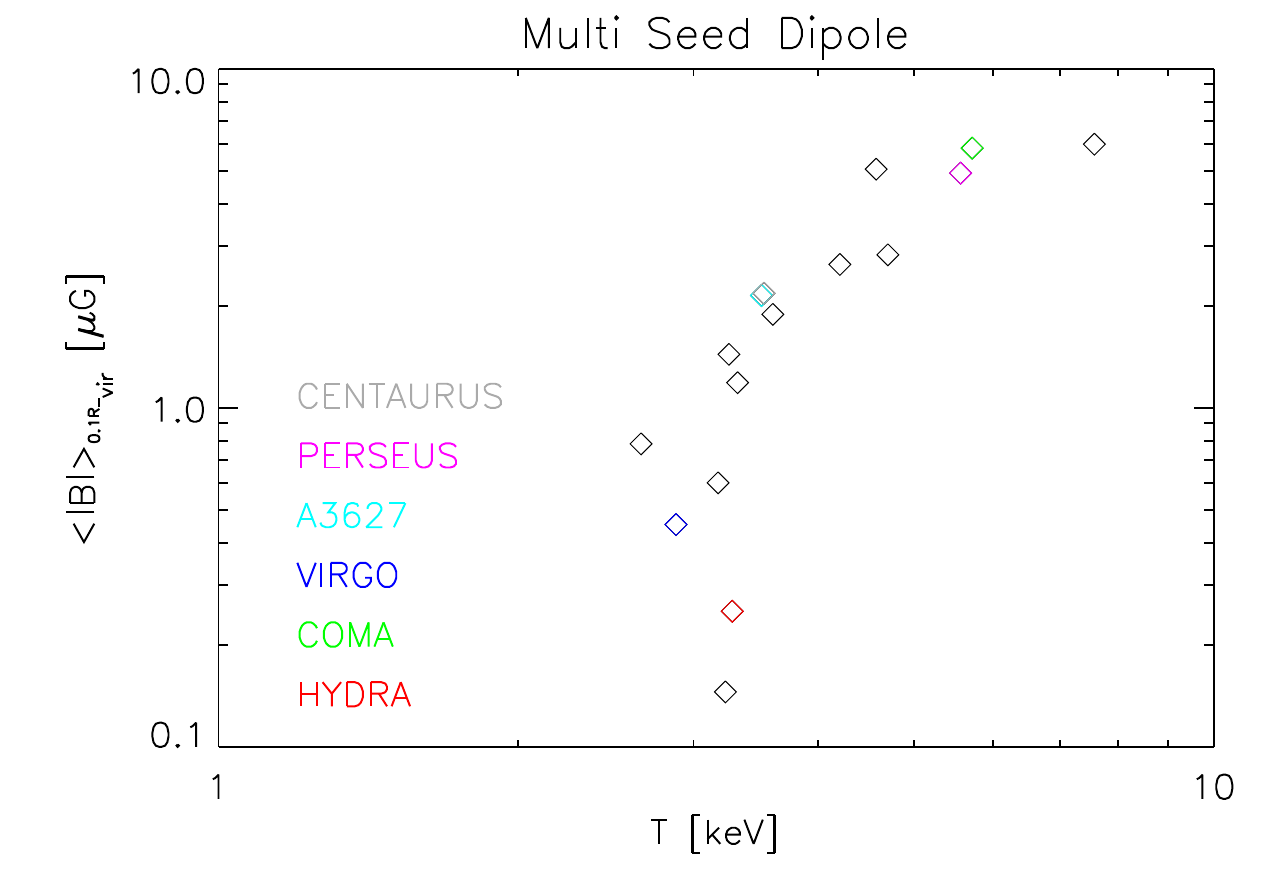}
    \end{minipage}
    \begin{minipage}[t]{0.45\textwidth}
    \includegraphics[width=1.0\textwidth]{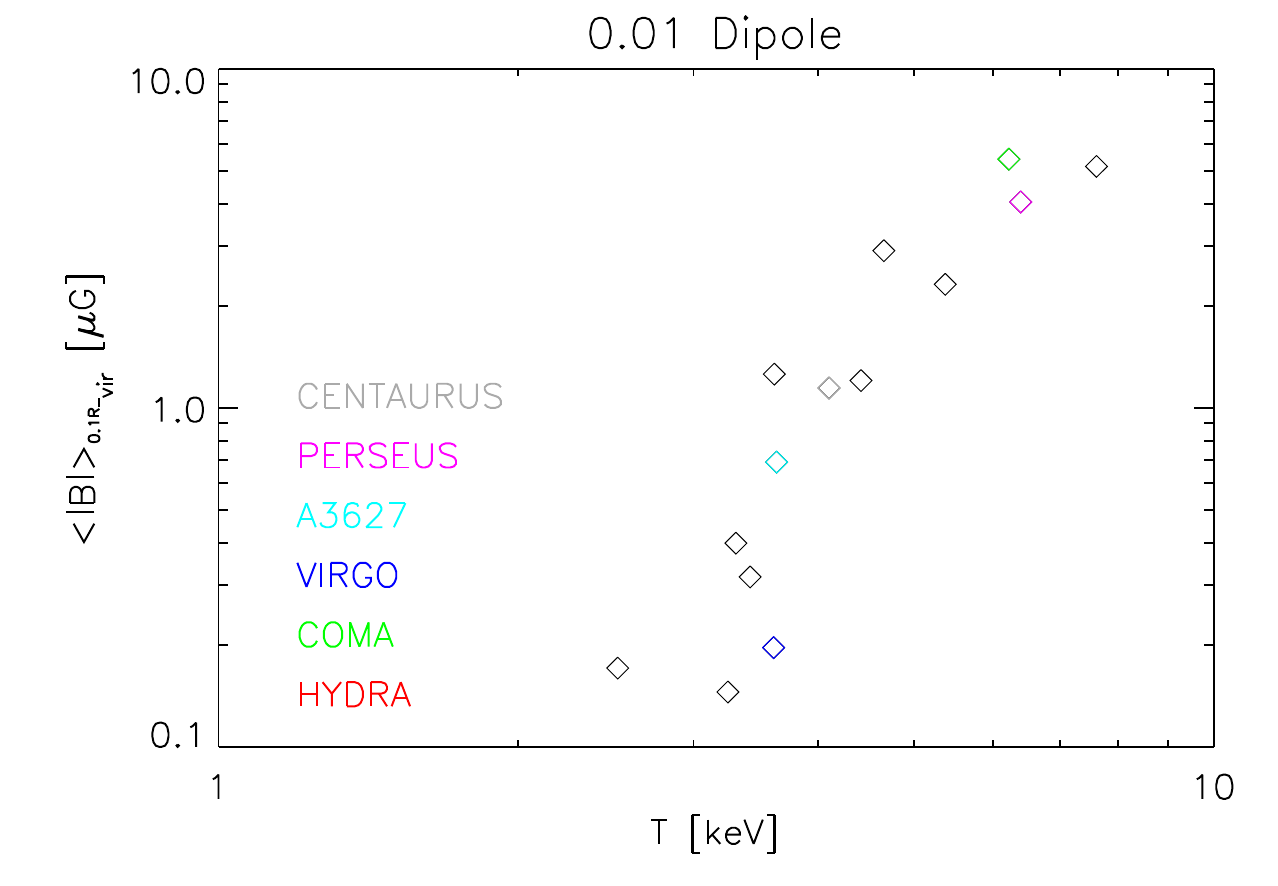}
    \end{minipage}
    \begin{minipage}[t]{0.45\textwidth}
    \includegraphics[width=1.0\textwidth]{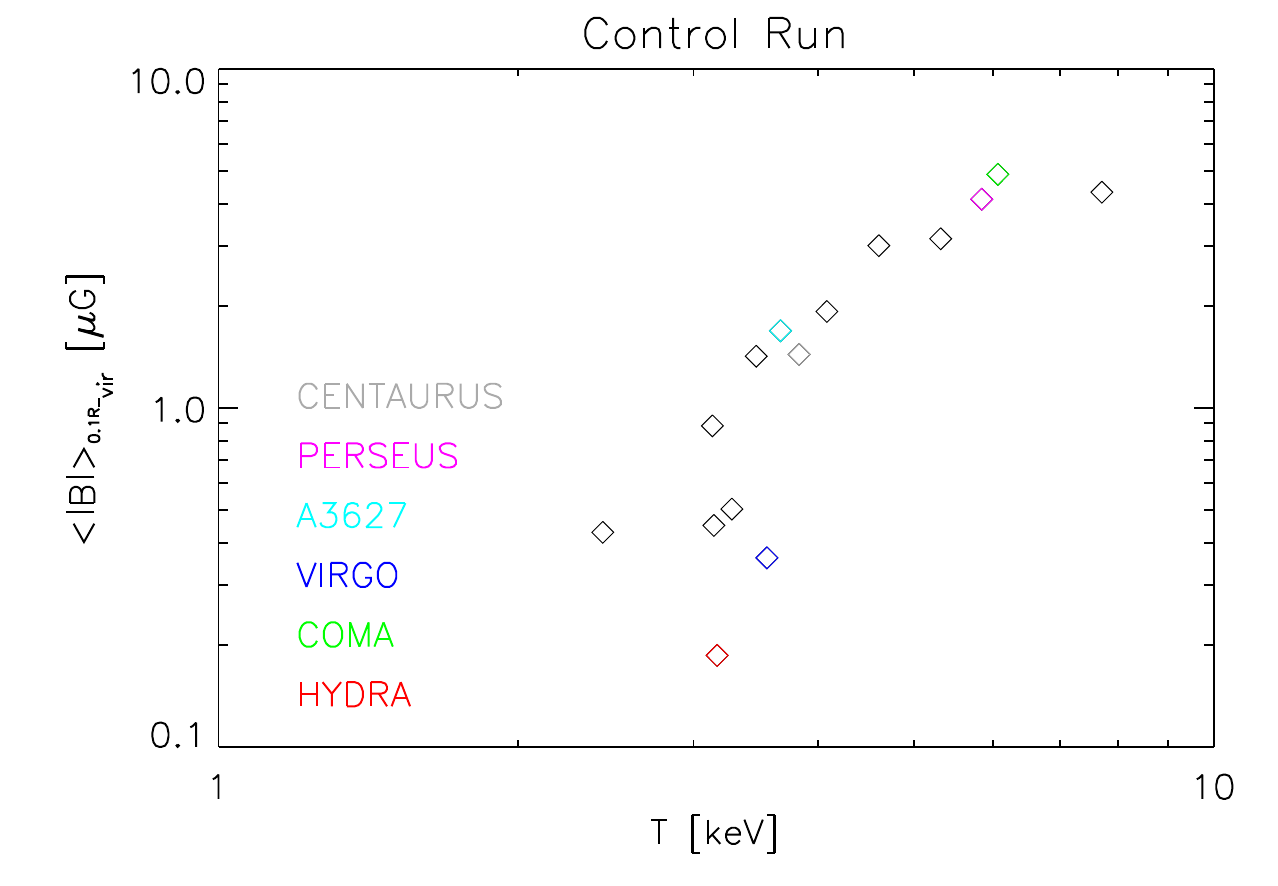}
    \end{minipage}
\caption{Same arrangement as figure \ref{b_prof} showing the mean
  magnetic field strength inside one tenth of the virial radius as
  function of the mean, mass-weighted temperature within the same
  radius. Clusters that have an observed counterpart are marked with
  different colors. }\label{btplots}
\end{figure*}

Figure \ref{btplots} shows the mean magnetic field within one tenth of
the virial radius as a function of mean, mass-weighted temperature for
the 16 most massive clusters from our simulations.  There is a strong
trend of the magnetic field with temperature (e.g. mass).  At low
temperatures (e.g. low masses) a large scatter is present, not only
for models with different seed fields, but also for clusters with
similar temperature. Simulations of low mass/temperature systems also
show a quite steep relation, indicating that for small systems, the
dynamical state is very important.  Here, the magnetic field
amplification is still in a regime, where individual merger events
strongly contribute to the amplification of the magnetic fields. At
intermediate and high temperatures the scatter reduces and the trend
becomes very regular.  For strong magnetic field models the mean field
flattens horizontally as a function of temperature, which indicates a
saturation of the magnetic field amplification. This can be seen well
when comparing runs with varying dipole strength (left column of
figure \ref{btplots}). Again, there is no significant dependence on
the original magnetic field structure assumed within the galactic
outflows as evident from comparing the {\it 0.1 Dipole } with the {\it
  Quadrupole} simulation. {\it Multi Seed} simulations only affect low
mass systems.

The normalisation of the average magnetic field strength might be
sensitive to the resolution of the simulation, as well as to the exact
details of saturation of the amplification mechanism.  This includes
the final amplitude of the mean magnetic fields as well as the
temperature (mass) scale at which clusters magnetic fields reach the
saturation regime.  The central magnetic fields (several $\mu$G) found
in our simulated clusters are in good agreement with observations
\citep[e.g. see values of][]
      {2008A&A...483..699G,2006A&A...460..425G,2004A&A...424..429M,bonafede08}.

\subsection{Synthetic rotation measurement profiles} 

\begin{figure*}
    \centering
    \begin{minipage}[t]{0.45\textwidth}
    \includegraphics[width=1.0\textwidth]{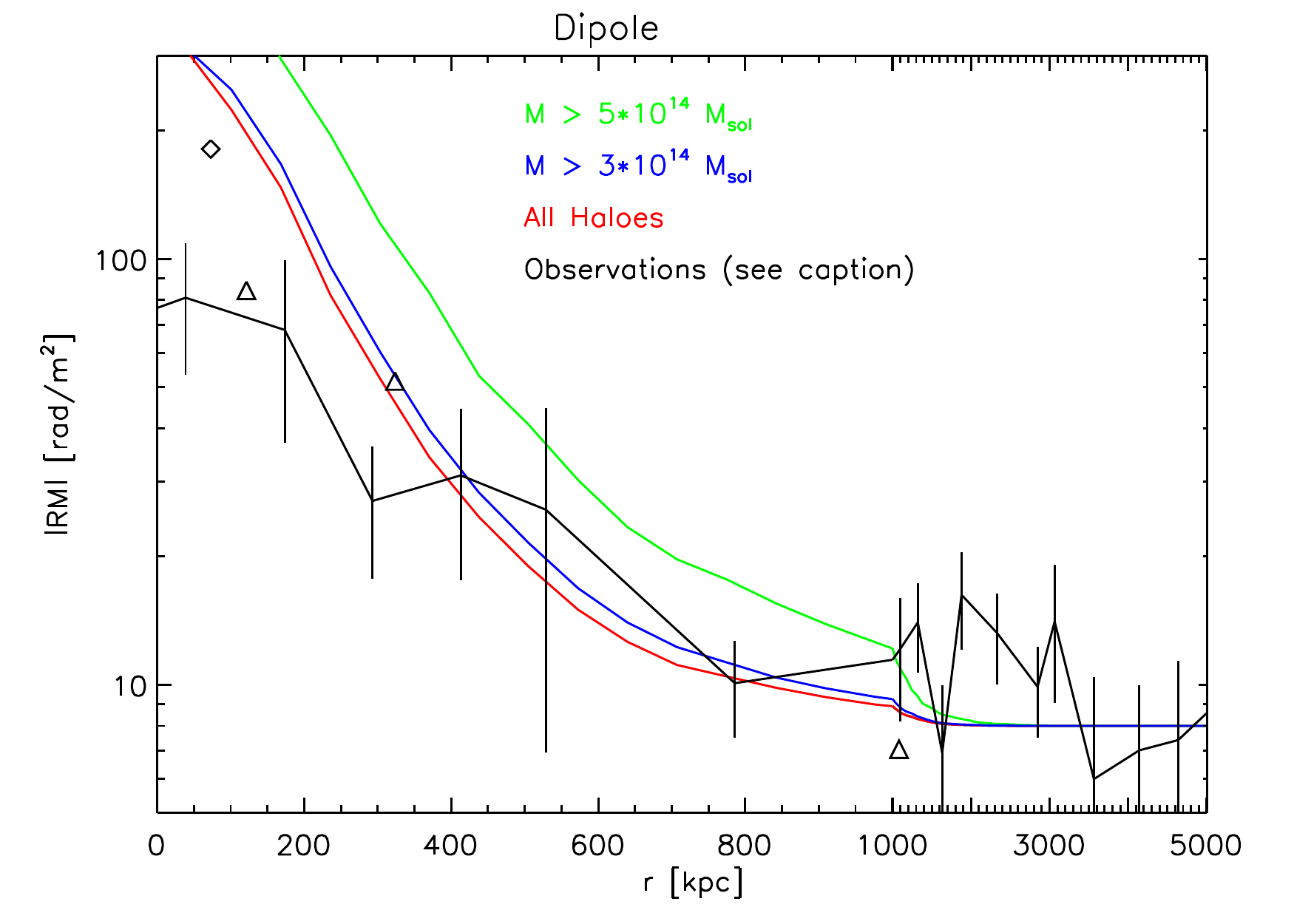}
    \end{minipage}
    \begin{minipage}[t]{0.45\textwidth}
    \includegraphics[width=1.0\textwidth]{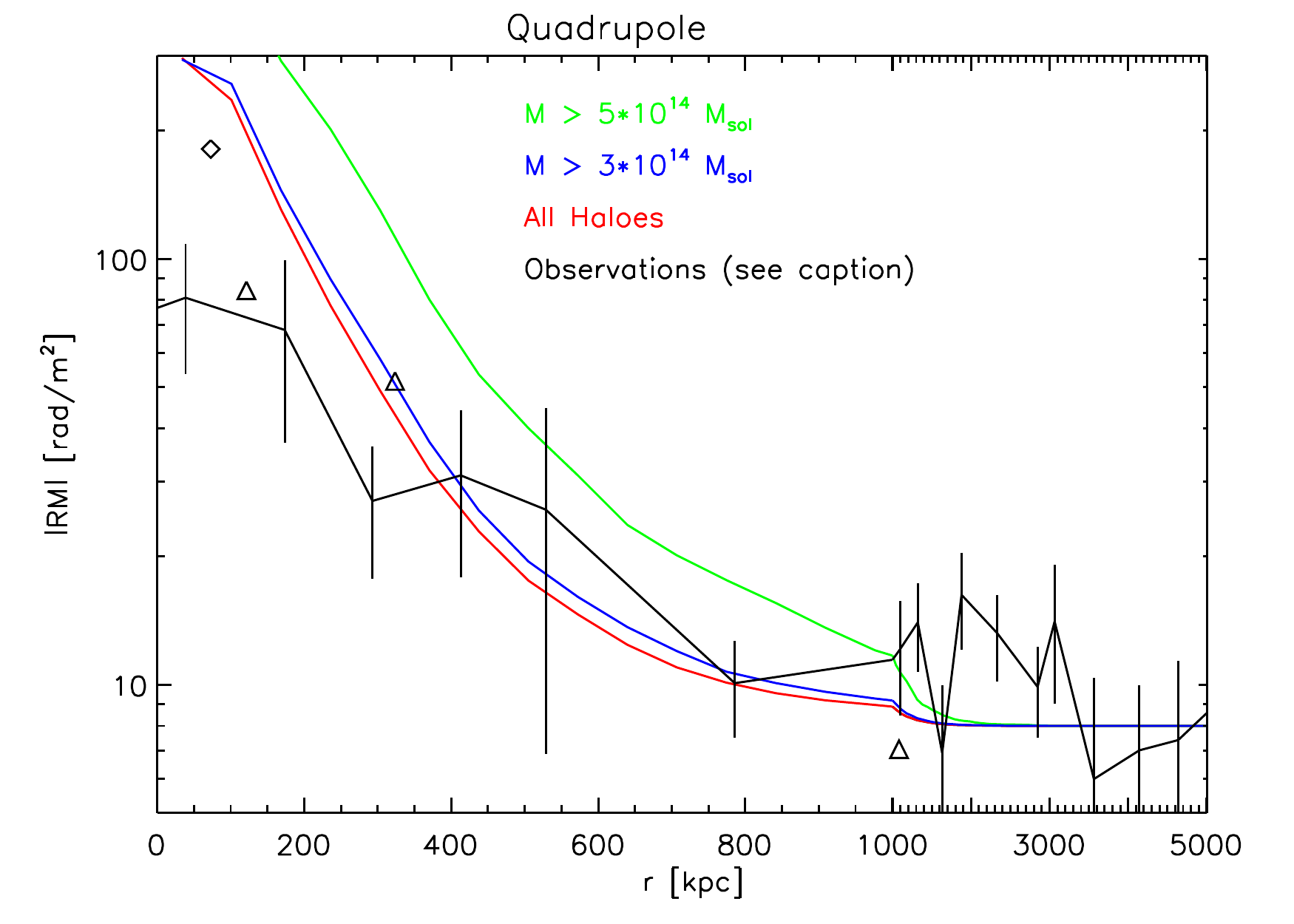}
    \end{minipage}
    \begin{minipage}[t]{0.45\textwidth}
    \includegraphics[width=1.0\textwidth]{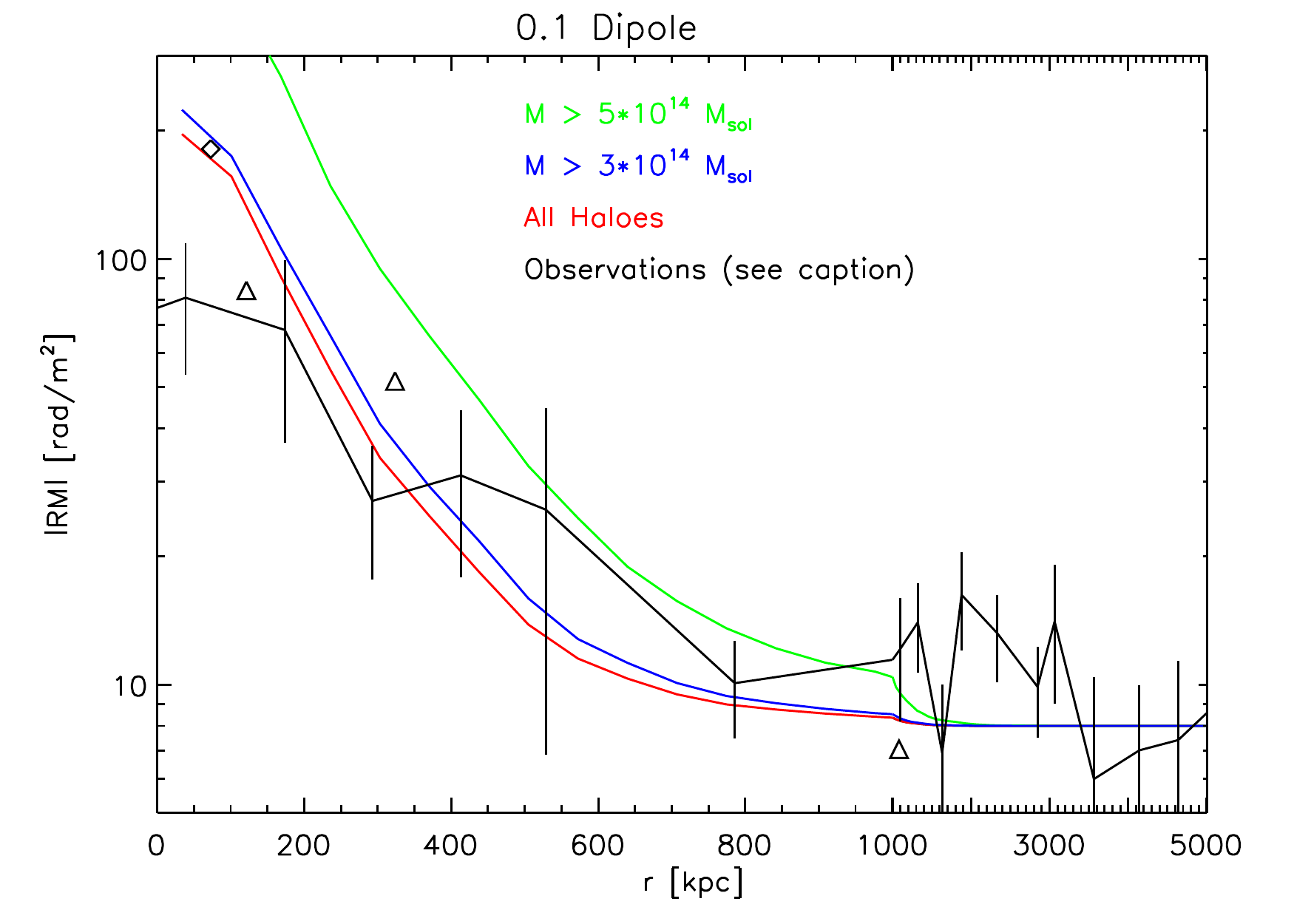}
    \end{minipage}
    \begin{minipage}[t]{0.45\textwidth}
    \includegraphics[width=1.0\textwidth]{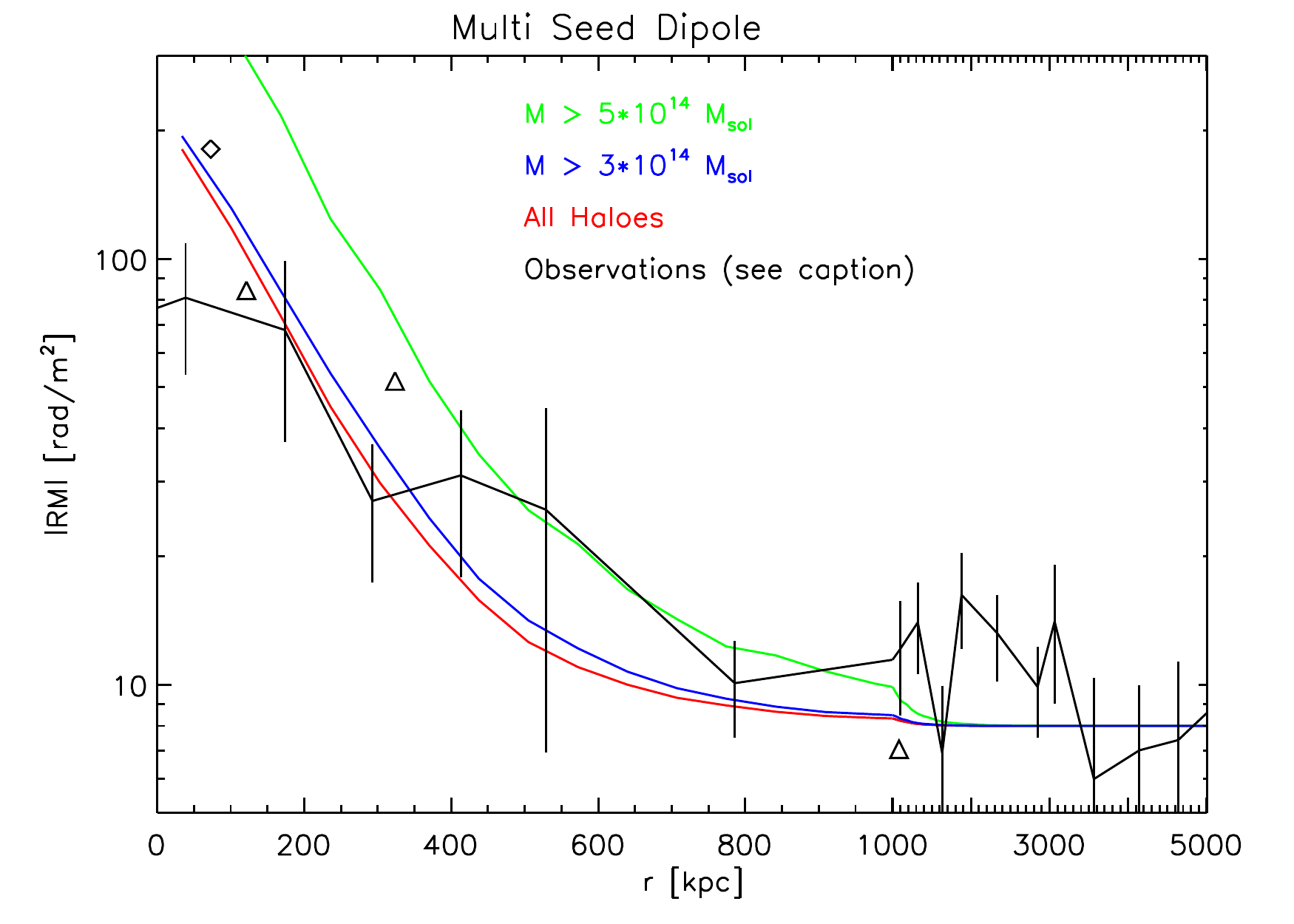}
    \end{minipage}
    \begin{minipage}[t]{0.45\textwidth}
    \includegraphics[width=1.0\textwidth]{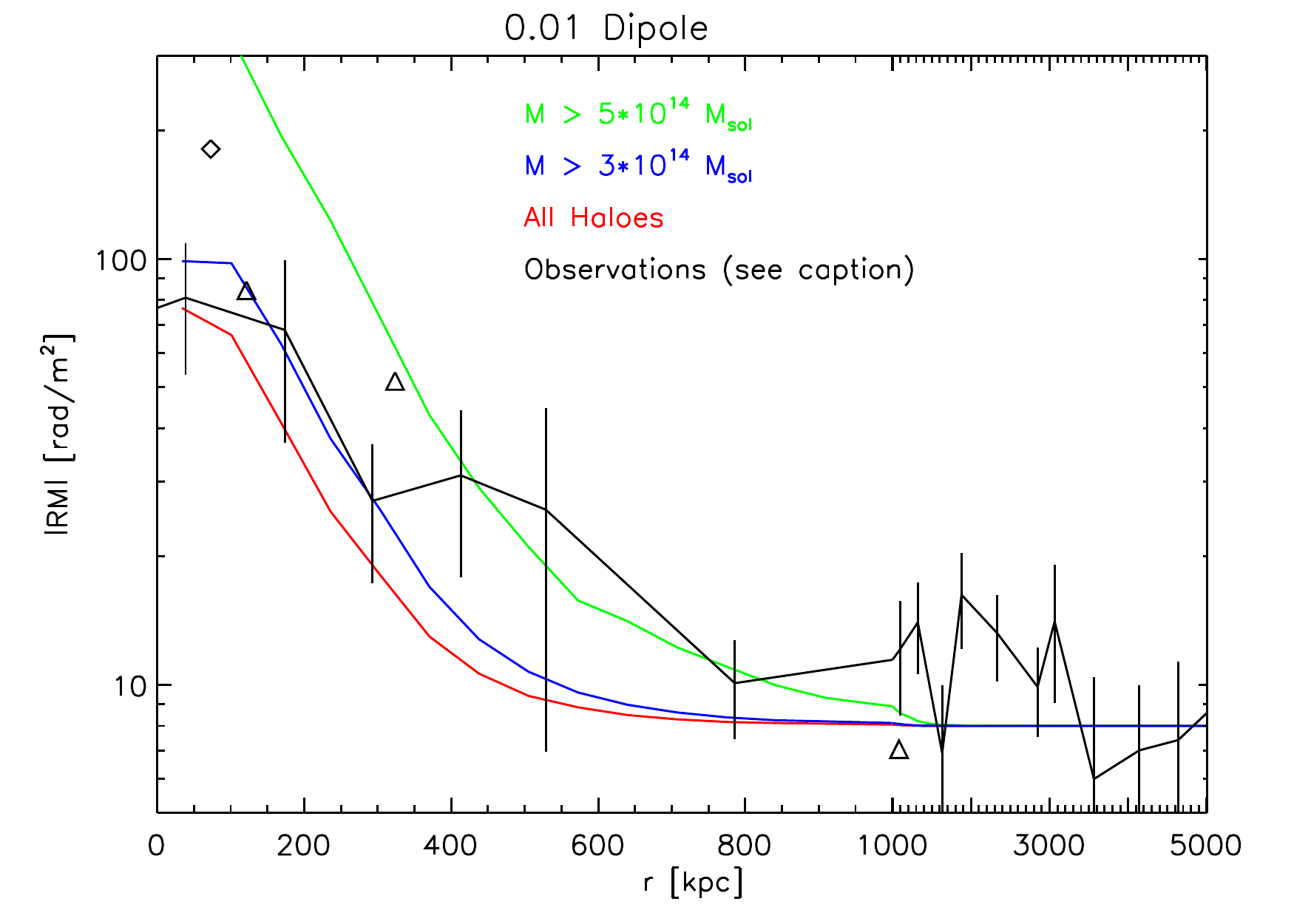}
    \end{minipage}
    \begin{minipage}[t]{0.45\textwidth}
    \includegraphics[width=1.0\textwidth]{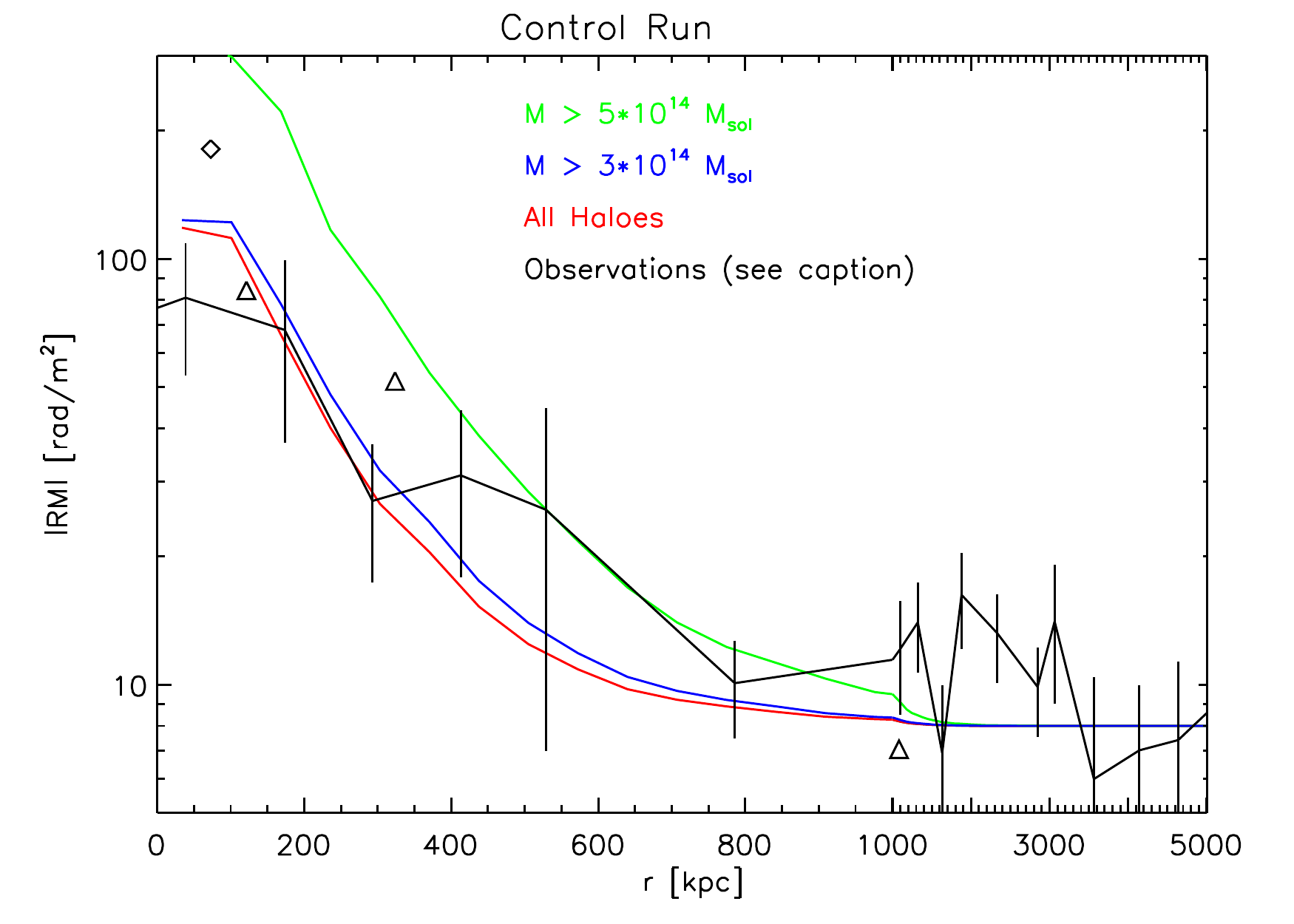}
    \end{minipage}
\caption{Same arrangement as figure \ref{b_prof} but showing the
  radial profiles of Faraday rotation. We binned the absolute value of
  the observed Faraday rotation as function of distance to the centres
  of the clusters and plotted the median forcing the radial intervals
  to contain always 15 data points. The data are obtained from
  combining three samples based on Abell clusters
  \citep{1991ApJ...379...80K,2001ApJ...547L.111C,2004astro.ph.11045J}.
  The error bars are obtained by bootstrapping the data points within
  each bin.  We also included in the plot the values inferred from
  three elongated sources (triangles) observed in the single galaxy
  cluster A119 \citep{1999A&A...344..472F} and one elongated source
  within the Coma cluster (diamond) \citep{1995AA...302..680F}.  For
  the simulations we build the median of the Faraday rotation for the
  16 most massive clusters (red line) and subsets restricting their
  mass to be larger than $3\times10^{14}M_\odot$ and
  $5\times10^{14}M_\odot$, blue and green line
  respectively.}\label{rmradprof}
\end{figure*}

To compare the simulated magnetic field from different seeds with
observations, we repeat the comparison with observational data as in
\citep{2002A&A...387..383D,2005JCAP...01..009D}.  Figure
\ref{rmradprof} shows a combination of three observational samples,
measuring the Faraday rotation of point like sources in or behind
Abell clusters as function of their distance to the center of the
clusters
\citep{1991ApJ...379...80K,2001ApJ...547L.111C,2004astro.ph.11045J}.
The black line shows the result obtained for the median of the
absolute values when radially binning using bins of 15 data points
each.  All three individual samples are statistically compatible with
each other and we combine all of them to allow a finer binning of the
data.  To calculate error bars we used the RMS of the median obtained
by bootstrapping the samples of each bin $1000$ times.  We over
plotted the values inferred from three elongated sources (triangles)
observed in the single galaxy cluster A119 \citep{1999A&A...344..472F}
and one elongated source within the Coma cluster (diamonds)
\citep{1995AA...302..680F}.  As discussed in previous work
\citep{2005JCAP...01..009D}, due to the construction of these
observational samples, the underlying selection function for the
contributing galaxy clusters is ill-defined.  This is especially
important, as the signal depends on the mass of the underlying galaxy
clusters, both due to the larger line of sight contribution from
massive systems as well as the predicted dependence of the magnetic
field strength on the cluster mass. One therefore expects the median
profile to depend on the mass function of the selected
clusters. However, as the cluster sample is composed of Abell
clusters, it is reasonable to assume that it mainly consists of
comparatively massive galaxy clusters. The additional data points
inferred from the two massive systems A119 and Coma (black symbols)
follow the curve from the combined sample (black line) resonable well,
indicating that the observed sample indeed compares well to the
predictions for more massive systems.

From the simulations we calculated synthetic Faraday rotation maps for
the 16 most massive clusters. Binning the individual maps in radial
bins we computed the median of the absolute value of the Faraday
rotation combining the same radial bins from all clusters. This
procedure was repeated for two subsets of the clusters, where we
restricted the sample to clusters with masses above
$3\times10^{14}M_\odot$ and $5\times10^{14}M_\odot$ respectively. The
results are shown as three colored lines in figure \ref{rmradprof}. In
general, the shape of the resulting radial profiles compares to
observations quite well; there is in particular no noticeable
difference between the {\it Control} run, which follows the evolution
of a primordial magnetic field and the {\it Dipole 0.1} run. The {\it
  Dipole 0.1} fits also best regarding the different strength for the
galactic halo field. In agreement with results presented before, there
is no visible difference between the quadrupole and the dipole
configurations for the magnetic field in the galactic winds. There are
also only very mild effects visible in the outer parts of the Faraday
rotation profiles for our run with the multiple seeding episode.

\begin{figure}
    \centering
    \begin{minipage}[t]{0.45\textwidth}
    \includegraphics[width=1.0\textwidth]{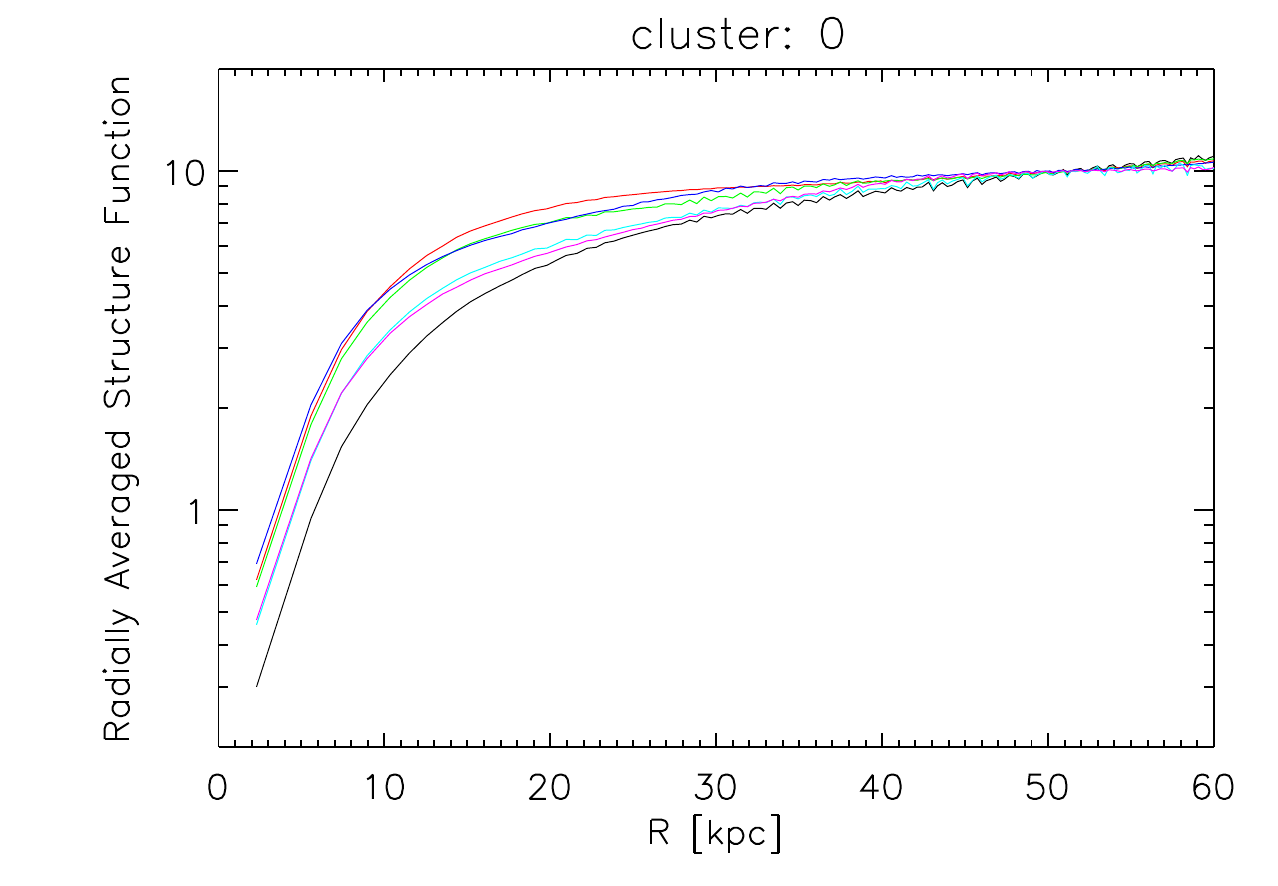}
    \end{minipage}
    \begin{minipage}[t]{0.45\textwidth}
    \includegraphics[width=1.0\textwidth]{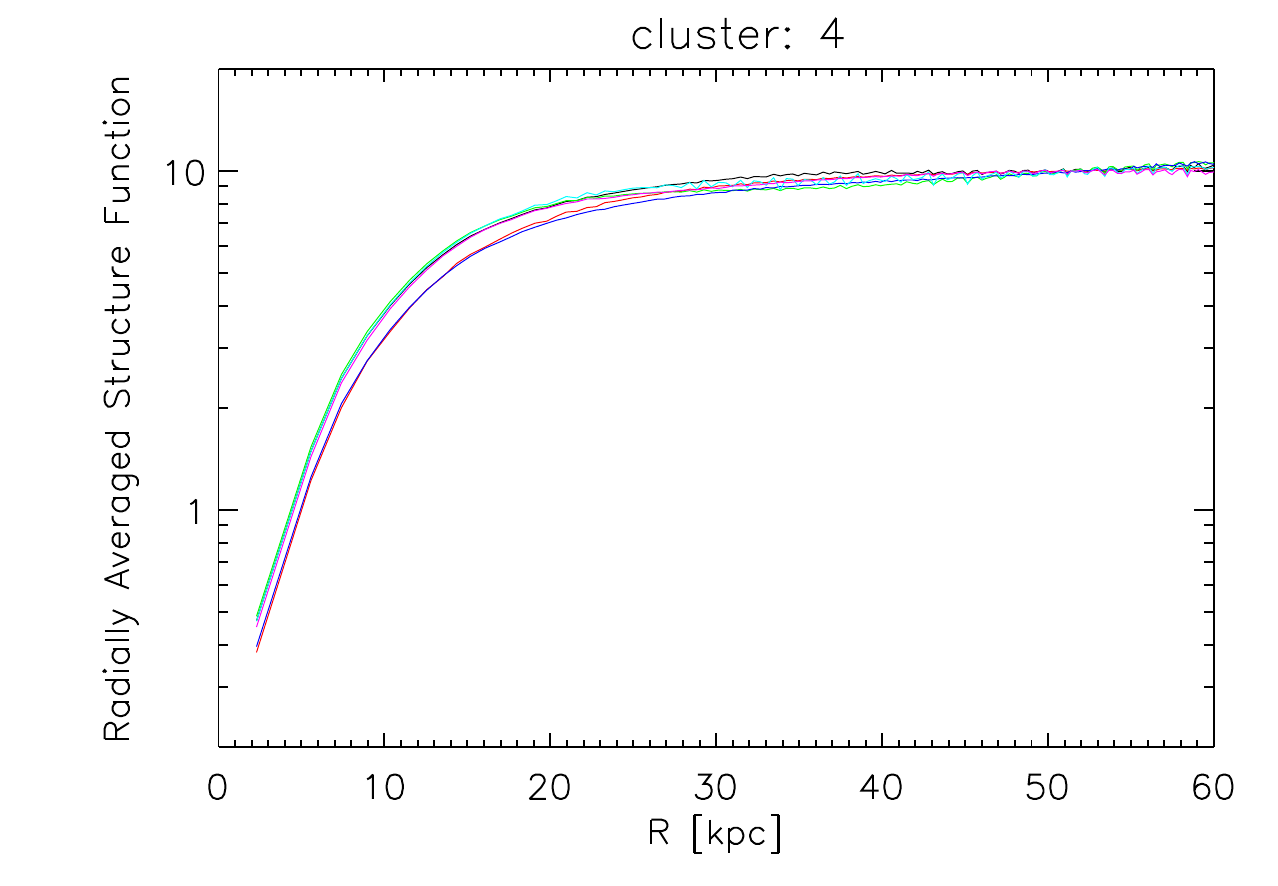}
    \end{minipage}
    \begin{minipage}[t]{0.45\textwidth}
    \includegraphics[width=1.0\textwidth]{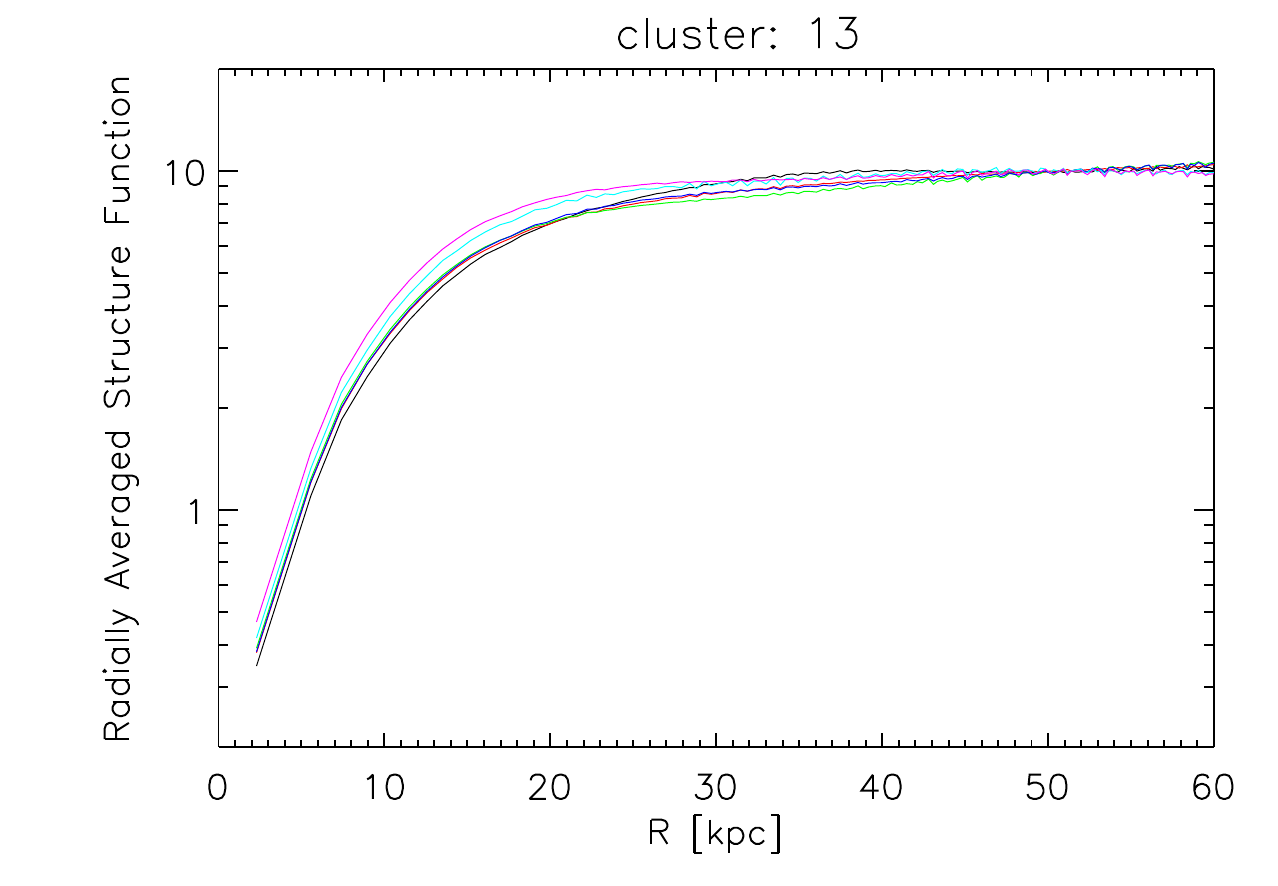}
    \end{minipage}
\caption{Shown is the structure function (for details see text) calculated 
from synthetic Faraday rotation maps obtained from different clusters 
displayed in the individual panels. The different lines correspond to  
different magnetic seed fields.}\label{strfunc}
\end{figure}

Finally, we calculated the projected structure function from some of
the synthetic Faraday rotation maps. This was done following
\citet{2004A&A...424..429M}, who calculated this for observed Faraday
rotation maps:
\begin{eqnarray}
S^{(1)}(dx,dy) &= \left<\left| RM(x,y) - RM(x+dx, y+dy)
\right|\right>,  
\end{eqnarray}
with $dx$ and $dy$ being the offsets from a pixel at position $(x,y)$.
The resulting matrix is then averaged in radial bins to finally obtain
the structure function. In figure \ref{strfunc} we show the structure
functions obtained from three of our clusters normalized at the
largest scales.  For the different models of the magnetic seed field
the structure functions only reveals some small differences in the
Faraday rotation of the first cluster. In general, there are no
significant indications of the seed magnetic field in the final
Faraday rotation structure, especially given that the effective
resolution limit of the simulation is about $15-20$ kpc in the central
parts of the clusters.

\section{Conclusions} \label{conclusions}

In this work we performed cosmological, magnetohydrodynamic
simulations following the evolution of magnetic fields on large scale
structures and in galaxy clusters. Coupling a semi-analytic model for
magnetized galactic winds as suggested by \citet{2006MNRAS.370..319B}
with our cosmological simulation we explored the possibility that the
magnetic fields in galaxy clusters originate from galactic outflows
during star-burst phases, further processed by structure formation.
We compared our results with the ones obtained by following a
primordial magnetic seed field \citep{2005JCAP...01..009D}.
Performing several simulations, we explored the effect of various
parameters of the adapted semi-analytic model relevant for the
strength of the magnetic seed field from the galactic outflows.  We
also explored the effect of the magnetic field configuration assumed
for the galactic outflows and of the seeding strategy. Our general
findings are:
\begin{itemize}
\item The typical magnetic field strengths of several $\mu$G in galaxy
  clusters as obtained from observations of Faraday rotation are well
  reproduced for a wide range of parameters of the galactic outflow
  model.
\item The general shape of the predicted Faraday rotation profile
  within clusters compares well with the sparse observational data
  available. Models that assume a field strength of $5\mu$G within the
  galactic halo reproduce the observed Faraday rotation profiles
  better than models with a ten times stronger or a ten times weaker
  halo magnetic field.
\item The properties of the final magnetic field in galaxy clusters do
  not depend on the exact field configuration within the magnetic
  outflows.  This confirms previous studies that the structure of the
  magnetic field in galaxy clusters is primarily driven by the
  velocity field induced by the structure formation process.
\item In massive galaxy clusters, the magnetic field amplification
  saturates around values of several $\mu$G. The mass (or temperature)
  scale on which this happens depends on the strength of the magnetic
  seed field, and probably also on the resolution of the simulation.
\item In systems where saturation effects start to play a significant
  role, we observe only a small scatter in the magnetic scaling
  relations, and in the shape of the radial, magnetic profiles.
\item In clusters where saturation effects are negligible the strength
  and configuration of the magnetic field strongly depends on the
  dynamic state. Therefore, we observe a large scatter in the magnetic
  scaling relations and in the shape of the radial, magnetic profiles.
\item Within galaxy clusters, the structures predicted from synthetic
  Faraday rotation maps do not depend significantly on the magnetic
  seed field, and agree well with observed ones.
\item In low density environments imprints of the magnetic seed fields
  are still present and can be observed, in principle. In these
  environments the galactic outflows forming at later epochs
  contribute significantly to the magnetic field configuration.
\end{itemize}

In summary, we confirm that galactic outflows and the subsequent
action of structure formation can explain the properties of the
magnetic fields observed in massive galaxy clusters. Their strength
and shape is attributable to the velocity field induced by structure
formation. We find that there are no measurable imprints of the
magnetic seed fields left in the synthetic Faraday rotation maps of
our simulated clusters. However, low density environments, like
filaments, still contain significant information on the magnetic seed
fields. Therefore, they may be used to discriminate proposed scenarios
once information on magnetic fields within these regions becomes
available with the next generation of radio instruments.

\section*{Acknowledgments}
We want to thank Caroline D'Angelo for carefully reading and 
improving the manuscript.

\appendix
\section{Conversion Formulae}\label{E2B}

\subsection{Dipole Energy}\label{m_Eb}

\begin{eqnarray}\label{FieldEnergy}
    E_{\mathrm{B}} &=& \frac{1}{8 \pi} \int  \vec{B}^{2} \,\mathrm{d}V \\
    \vec{B} &=&\frac{1}{4\pi} \frac{3 \vec{n} \left( \vec{n} \cdot \vec{m}\right) - \vec{m}}{\left| \vec{r} \right|}
\end{eqnarray}

To calculate the magnetic moment $\vec{m}$ of a dipole of magnetic
energy $E_{\mathrm{B}}$, one assumes, that $\vec{m} ||
\vec{e}_{\mathrm{z}}$, as the moment can be aligned into any other
direction by a simple rotation of the coordinate system. The unit
vector $\vec{n}$ then becomes $ z / \left| z \right| $. \\ 
Since the field
$\vec{B}$ diverges for $r \rightarrow 0$ we use a softening length of
$\epsilon = 14$, which corresponds to the SPH smoothing length. The
outer boundary $R_{\mathrm{max}}$ takes the finite size of the halo
into account. Setting $\mu_{0} = 1$ it follows:
\begin{eqnarray*}
E_{\mathrm{B}}  &=& \frac{1}{8\pi} \int \frac{1}{(4\pi)^{2}}
\frac{\left(3 \vec{n} \left( \vec{n} \cdot \vec{m} \right) -\vec{m}
  \right)^{2}}{\left(\left| \vec{r}\right|^{3}
  +\epsilon^{3}\right)^{2}}  \,\mathrm{d}V \\ &=&  \frac{1
}{120\pi^{3}} \int \frac{\mathrm{d}V}{\left(\left| \vec{r}
  \right|^{3}+\epsilon^{3}\right)^{2}} \cdot \left( 3 \left(
\frac{z}{\left| \vec{r} \right|} m_{\mathrm{z}} \right)^{2}  +
m_{\mathrm{z}}^{2} \right) \\ &=&
\frac{m_{\mathrm{z}}^{2}}{4\left(4\pi \right)^{2}} \cdot
\int\limits_{0}^{R_{\mathrm{max}}} \int\limits_{0}^{\pi} \frac{3
  r^{2}}{\left( r^{3} + \epsilon^{3} \right)^{2}} \cdot \sin\phi \,
\cos^{2}\phi \,\mathrm{d}r \, \mathrm{d}\phi + \nonumber \\ & &
\frac{m_{\mathrm{z}}^{2}}{4\left(4\pi \right)^{2}}
\cdot\int\limits_{0}^{R_{\mathrm{max}}} \int\limits_{0}^{\pi}
\frac{r^{2}}{\left( r^{3} + \epsilon^{3} \right)^{2}} \cdot \sin\phi
\, \mathrm{d}r\, \mathrm{d}\phi ,
\end{eqnarray*}
where we used spherical coordinates. It follows
\begin{eqnarray}
    m_{\mathrm{z}} &=& \sqrt{48 \pi^{2} E_{\mathrm{B}} \cdot \left(
      \frac{1}{\epsilon^{3}} -
      \frac{1}{R^{3}_{\mathrm{max}}+\epsilon^{3}}  \right)^{-1} }.
\end{eqnarray}

\subsection{Quadrupole Energy}\label{qm_Eb}
We constitute the quadrupole as a superposition of 2 dipoles at a
distance $d$, with dipole moments $\vec{m}_{1} = m_{\mathrm{z}}
\vec{s}$ and  $\vec{m}_{2} = m_{\mathrm{z}} \left(- \vec{s}\right)$,
``pointing'' into opposite directions. Introducing a softening length
$\epsilon$, the field is given by 
\begin{eqnarray*}
     \vec{B} &=& \vec{B}_{+}^{\mathrm{DIP}} +
     \vec{B}_{-}^{\mathrm{DIP}}
\end{eqnarray*}
As the field energy does not depend on the direction of the dipole
\begin{eqnarray*}
    E_{\mathrm{B}} &=& \frac{2}{8\pi} \int \vec{B}^{2}\, \mathrm{d}V
    + \frac{2}{8\pi}  \int \vec{B}_{+} \vec{B}_{-}\, \mathrm{d}V.
\end{eqnarray*}
We define 
\begin{eqnarray*}
    \vec{r}_{\pm} &=& \vec{r} \mp \frac{d}{2}\, \vec{s},  \\
	 r_{\pm} &=& \left| \vec{r}_{\pm} \right|, \\
	 \vec{n}_{\pm} &=& \frac{\vec{r}_{\pm}}{r_{\pm}}, \\ 
	 n_{\pm}^{\mathrm{z}} &=& \vec{n}_{\pm} \cdot \vec{e}_{\mathrm{z}} ,\\ 
	 \vec{m}_{\pm} &=& \pm m_{\mathrm{z}} \vec{s},
\end{eqnarray*}
and the field energy becomes
\begin{eqnarray*}
    E_{\mathrm{B}} &=& \frac{2 m_{\mathrm{z}}^{2}}{48\pi^{2}} \alpha
    + \frac{-2 m_{\mathrm{z}}^{2}}{8\pi (4\pi)^{2}}\cdot \gamma,
\end{eqnarray*}
where
\begin{eqnarray*}
    \alpha &=& \left( \frac{1}{\epsilon^{3}} -
    \frac{1}{R_{\mathrm{max}}^{3}+\epsilon^{3}} \right),
\end{eqnarray*}
and
\begin{eqnarray*}
  \gamma &=&
    \int   \frac{9\left(\vec{n}_{+} \vec{n}_{-} \right)
      n_{+}^{\mathrm{z}} n_{-}^{z} - 3 (\vec{n}^{\mathrm{z}}_{+})^{2}
      - 3 (\vec{n}^{\mathrm{z}}_{-})^{2}+1}{\left(r_{+}^{3}+
      \epsilon^{3}\right)\left(r_{-}^{3}+ \epsilon^{3}\right)}\, dV. \\
\end{eqnarray*}

\bibliographystyle{mn2e} \bibliography{julius,master3}

\label{lastpage}
\end{document}